 SP 2008/0/3/01

\documentclass[preprint,12pt]{elsarticle}

\usepackage{amssymb}
\usepackage{graphicx}
\usepackage{textcomp}
\usepackage{listings}
\usepackage{multibib}

\begin{document}

\begin{frontmatter}



\title{Smart Contracts Software Metrics: a First Study}


 \author{R. Tonelli$^a$, G. Destefanis$^b$, M. Marchesi$^a$, M. Ortu$^c$}

\address{Dept. of Matehmatics and Informatics, University of Cagliari, Italy $^a$ \\
roberto.tonelli@dsf.unica.it - michele@diee.unica.it\\
School of Computer Science, University of Hertfordshire, United Kingdoms $^b$\\
g.destefanis@herts.ac.uk\\
Dept. of Electric and Electronic Engineering, University of Cagliari, Italy $^c$\\
marco.ortu@diee.unica.it
}

\begin{abstract}
Smart contracts (SC) are software codes which reside and run over a blockchain.
The code can be written in different languages with the common purpose 
of implementing various kinds of transactions onto the hosting blockchain, 
They are ruled by the blockchain infrastructure and work in order
to satisfy conditions typical of traditional contracts. 
The software code must satisfy constrains strongly context dependent 
which are quite different from traditional software code. 
In particular, since the bytecode is uploaded in the hosting blockchain, 
size, computational resources, interaction between 
different parts of sofware are all limited and even if the specific 
software languages implement more or less the same constructs of 
traditional languages there is not the same freedom 
as in normal software development. 
SC software is expected to reflect these constrains on SC software metrics 
which should display metric values characteristic of the domain 
and different from more traditional software metrics. 
We tested this hypotesis on the code of more than twelve thousands SC 
written in Solidity and uploaded on the Ethereum blockchain.
We downloaded the SC from a public repository and computed the statistics of 
a set of software metrics related to SC and compared them to 
the metrics extracted from more traditional software projects. 
Our results show that generally Smart Contracts metrics have ranges more restricted 
than the corresponding metrics in traditional software systems. 
Some of the stylized facts, like power law in the tail of the distribution of some metrics,
are only approximate but the lines of code follow a log normal distribution 
which reminds of the same behavior already found in traditional software systems.

\end{abstract}

\end{frontmatter}

\section{\label{sec:level1}Introduction\protect}

Smart Contracts have gained tremendous popularity in the past few years, to the point that billions of US Dollars are currently exchanged every day through such technology. However, since the release of  the Frontier network of Ethereum in 2015, there have been  many cases in which the execution of Smart Contracts managing Ether coins lead to problems or conflicts.
Smart Contracts rely on a non-standard software life-cycle, according to which, for instance,  delivered applications can hardly be updated or bugs resolved by releasing a new version of the software.
Furthermore their code must satisfy to constraints typical of the domain, like the following: they must be light, the deployment on the blockchain must 
take into account the cost in terms of some criptovalue, their operational cost, againg in terms of criptovalue, must be limited, they are 
immutable, since the bytecode is inserted into a blockchain block once and forever.

The idea of Sc was originally described by cryptographer Nick Szabo in 1997, 
as a kind of digital vending machine. In his paper \cite{Szabo}, 
he imagined how users could input data or value, and receive a finite item from a machine.

More in general, {\em smart contracts} are self-enforcing agreements, i.e. contracts, implemended through a computer program whose execution enforces the terms of the contract. The idea is to get rid of a central control authority, entity or organization which both parties must trust and delegate such role to the {\em correct} execution of a computer program. Such scheme can thus rely on a decentralized system automatically  managed by machines. 
The Blockchain technology is the instrument for delivering the trust model envisaged by smart contracts.

Since smart contracts are stored on a blockchain, they are public and transparent, immutable and decentralised, and since blockchain resources 
are costly, their code size cannot exceed domain specific constrains. 
Immutability means that when a smart contract is created, it cannot be changed again.

Smart contracts can be applied to many different scenarios: banks could use them to issue loans or to offer automatic payments; insurance companies could use them to automatically process claims according to agreed terms; postal companies for payments on delivery.

A \textit{Smart Contract} (SC) is a \textit{full-fledged program} stored in a blockchain by a \textit{contract-creation} transaction. A SC is identified by a \textit{contract address} generated upon a success creation transaction. A blockchain state is therefore a mapping from addresses to accounts. 
Each SC account holds an {\em amount of virtual coins} (Ether in our case), and has its own private {\em state} and  {\em storage}. 

The up-to-date SC progrmmaing language is Solidity which runs on the Ethereum Virtual Machine (EVM) on the Ethereum blockchain. 
Since this is currently the most popular paradigm, we focus our attention on Solidity.  
An Ethereum SC account hence typically holds its executable code and a state consisting of: 
\begin{itemize}
\item a private storage
\item the amount of virtual coins (Ether) it holds, i.e. the contract {\em balance}.
\end{itemize}

Users can transfer Ether coins using transactions, like in Bitcoin, and additionally can \textit{invoke} contracts using
\textit{contract-invoking} transactions. Conceptually, Ethereum can be viewed as a huge \textit{transaction-based state machine}, where its state is updated after every transaction and stored in the blockchain.

Smart Contracts source code manipulate variables in the same way as traditional imperative programs. 
At the lowest level the code of an Ethereum SC is a stack-based bytecode language run by an Ethereum virtual machine (EVM) in each node. 
SC developers define contracts using high-level programming languages. One such language for Ethereum is Solidity \cite{solidity2014} (a JavaScript-like language), which is compiled into EVM bytecode. 
Once a SC is created at an address \textit{X}, it is possible to invoke it by sending a contract-invoking transaction to the address \textit{X}.
A contract-invoking transaction typically includes: 
\begin{itemize} 
\item payment (to the contract) for the execution (in Ether).
\item input data for the invocation.
\end{itemize}

\subsubsection{Working Example}
\label{subsubsec:ethereum-example}

Figure \ref{fig:smart-contract-code} shows a simple example of SC reported in \cite{Luu}, which rewards anyone who solves a problem and submit the solution to the SC.

\begin{figure}
\label{fig:SC}
\begin{lstlisting}[][frame=single]  

contract Puzzle {
 
  	 address public owner ;
    bool public locked ;
    uint public reward ;
    bytes32 public diff ;
    bytes public solution ;
    
    function Puzzle ()  {// constructor
	 	owner = msg.sender ;
     	reward = msg.value ;
      	locked = false ;
      	diff = bytes32 (11111); // pre-defined difficulty
	}
    
 	function (){ // main code , runs at every invocation
		if ( msg.sender == owner ){ // update reward
			if ( locked )
				throw ;
        	owner.send(reward);
          reward = msg.value ;
    	} else if ( msg.data.length > 0){
        	// submit a solution
    		if ( locked ) throw ;
    		if ( sha256 ( msg.data ) < diff ){
    			msg.sender.send(reward); // send reward
    			solution = msg.data ;
    			locked = true ;
 			}
		}
	}
}

\end{lstlisting}
\caption{\label{fig:smart-contract-code} Smart Contracts example.}
\end{figure}

A \textit{contract-creation} transaction containing the EVM bytecode for the contract in Figure~\ref{fig:smart-contract-code} is sent to miners. Eventually, the transaction will be accepted in a block, and all miners will update their local copy of the blockchain: first a \textit{unique} address for the contract is generated in the block, then each miner executes locally the \textit{constructor} of the \textbf{Puzzle} contract, and a local storage is allocated in the blockchain. Finally the EVM bytecode of the anonymous function of \textbf{Puzzle} (Lines 16+) is added to the storage.

When a \textit{contract-invoking} transaction is sent to the address of \textbf{Puzzle}, the function defined at Line 16 is executed by default.
All information about the sender, the amount of Ether sent to the contract, and the input data of the invoking transaction are stored in a default input variable called \textit{msg}. 
In this example, the \textit{owner} (namely the user that created the contract) can update the \textit{reward} (Line 21) by sending Ether coins stored in \texttt{msg.value} ({\tt if} statement at Line 17), after sending back the current \textit{reward} to the \textit{owner} (Line 20).

In the same way, any other user can submit a solution to \textbf{Puzzle} by a \textit{contract-invoking} transaction with a \textit{payload} (i.e., \texttt{msg.data}) to claim the reward (Lines 22-29). 
When a correct solution is submitted, the contract sends the reward to the sender (Line 26).

\subsubsection{Gas system}
It is worth remarking that a SC is run on the blockchain by each miner deterministically replicating the execution of the SC bytecode on the local copy of the blockchain. This, for instance, implies that in order to guarantee coherence across the copies of the blockchain, code must be executed in a strictly deterministic way (and therefore, for instance, the generation of random numbers may be problematic).

Solidity, and in general high-level SC languages, are Turing complete in Ethereum. In a decentralised blockchain architecture Turing completeness may be problematic, e.g. the replicated execution of infinite loops may potentially \textit{freeze} the whole network.

To ensure fair compensation for expended computation efforts and limit the use of resources, Ethereum pays miners some fees, proportionally to the required computation.
Specifically, each instruction in the Ethereum bytecode requires
a pre-specified amount of {\em gas} (paid in Ether coins). 
When users send a \textit{contract-invoking} transaction, they  must specify the amount of gas provided for the execution, called \textit{gasLimit}, as well as the price for each gas unit called \textit{gasPrice}. A miner who includes the transaction in his proposed block receives the transaction fee corresponding to the amount of gas that the execution has actually burned, multiplied by \textit{gasPrice}. If some execution requires more gas than
\textit{gasLimit}, the execution terminates with an exception, and the state is rolled back to the initial state of the execution. In this case the user pays all the \textit{gasLimit} to the miner as a counter-measure against resource-exhausting attacks \cite{luu2015demystifying}.

The code in Fig. \ref{fig:SC} displays typical features of the Solidity SCs code: the \textit{Contract} declaration, addresses declarations and mapping, 
owner data managing and the functions with the specific code for implementing the contract and transactions between blockchain addresses. 
Most of the control structures from JavaScript are available in Solidity except for switch and goto. So there is: if, else, while, do, for, break, continue, return, ? :, with the usual semantics known from C or JavaScript.

Functions of the current contract can be called directly (Internal Function Calls), 
also recursively. These function calls are translated into simple jumps inside the EVM. 
This has the effect that the current memory is not cleared, i.e. passing memory references to internally-called functions is very efficient. 
Only functions of the same contract can be called internally. 
The expressions this.g(); and c.g(); (where c is a contract instance) are also valid function calls, 
but this time, the function will be called as External Function Call, 
via a message call and not directly via jumps. 
Functions of other contracts have to be called externally. For an external call, all function arguments have to be copied to memory.
When calling functions of other contracts, the amount of criptocurrency (Wei) sent with the call and the 
gas can be specified with special options .value() and .gas(), respectively.
Inheritance between contracts is also supported. Being the format inspired to classes of object oriented programming languages, 
it is straighforward to define and compute some of the software metrics typically encountered in object oriented software systems, like 
number of lines of code, comments, number of methods or functions, cyclomatic complexity and so on, while it is somehow more difficult 
to recognize software metrics related to communication between smart contracts, since these can be ruled by blockchain transactions among 
contracts, which can act somehow as code libraries. 

On the other hand smart contracts are deployed and work on the blockchain infrastructure and it is thus likely that typical value 
of the same metrics can differ from the typical values of the same metrics in traditional software systems. 

It became thus interesting, even from a software engineering point of view, to perform a statistical analisys of SCs software metrics 
and to compare the data with those diplayed by traditional software systems. It is also of primary interest 
to examine the connnection between software metrics and software quality, a field of research well established in traditional 
software, in the specific domain of smart conracts given that it is well known that SCs code vulnerability have been exploited to 
stole value in criptocurrencies from smart contratcs. 

In this paper we perform the analysis on a data set of 12094 smart contracts downloaded 
from etherscan.io, a platform allowing enhanced browsing of ethereum blockchain and smart contracts, up to the January 24 2018. 
We collected the blockchain addresses, the Solidity source code, the ABI and the bytecode of each contract 
and extracted a set of standard and SC-specific software metrics such as number of lines of smart contract code (LOCSC), 
line of comments, blank lines, number of functions, cyclomatic complexity, number of events calls, number of mappings 
to addresses, number of payable, number of modifiable and so on. 
We analyzed the statistical distributions underlying such metrics to discover if they exhibit the same statistical properties 
typical of standard software systems or if the SM costraints act so that a sensible variation in these distribution 
can be detected. Furhtermore we devise a path to the analysis of which and to what extent the SC metrics influence 
samrt contract performance, usage in the blockchain, vulnerabilities, and possible other factors related to the specific 
contracts which can be reflected on the domain of application for which the smart contract has been deployed, 
like, for example, to implement and rule an initial coin offer (ICO), to control a chain of certification like in medical 
applications and so on.


\section{Related works}\label{sec:relatedWorks}

Blockchain technology and Smart Contracts rised an exponentially increasing interest in the last years in different fields of research. 
Organizations such as banking and financial institutions, and public and regulatory bodies, started to explicitly talk of the importance 
of these new technologies.Software Engineering specific for blockchain applications and Smart Contract is still in its infancy \cite{BOSE} 
and in particular the investigation of the relationships among Smart Contracts Software Metrics (SCSM) and code quality, 
SC performaces, vulnerability, maintainability and other software features is completely lacking.
Smart Contracts and blockchain have been discussed in many textbooks \cite{MelanieSwan} and documents over the internet, where 
white papers usually cover the specific topic of interest \cite{white1,white2,white3}.

Ethereum defines a smart contract as a transaction protocol that executes the terms of a contract
or group of contracts on a cryptographic blockchain \cite{MS65}. SC operate autonomously with no 
entity controlling the majority of its tokens, and its data and records of operation must be 
cryptographically stored in a public, decentralized blockchain \cite{MelanieSwan}.

Smart Contract vulnerabilities have been analyzed in \cite{smart1}, \cite{smart2}, \cite{smart3}. 
A taxonomy of Smart Contract is performed in \cite{Bartoletti:2017}, where Smart Contracts are classified according 
to their purpose. These are divided into wallets, financial, notary, game and library. 

Authors in \cite{Luu} investigate the security of running smart contracts based on Ethereum in an open distributed network like those of cryptocurrencies and 
introduce several new security problems in which an adversary can manipulate smart contract execution to gain profit.

Obviously Smart Contract scientific literature is quite limited due to their recent creation.
On the other hand there is a pletora of results and information to rely on produced in the last decades for what concernes 
the relationship among software metrics and software quality, maintainability, reliability, performance defectiveness 
and so on. 
Measuring software to get information about its properties and quality is one of the main issues in modern software engineering. 
Limiting ourselves to object-oriented (OO) software, one of the first works dealing with this problem is the one 
by Chidamber and Kemerer (CK) \cite{Chidamber:91}, who introduced the popular CK metrics suite for OO software systems \cite{CK:1994}. 
In fact, different empirical studies showed significant correlations between some of CK metrics and bug-proneness 
\cite{CK:1998} \cite{Basili:1996} \cite{Subramanyam:2003} \cite{Gyimoothy:2005}, \cite{MOOD}.
Metrics have been defined also on software graphs and were found most correlated to software quality \cite{Zimmermann:2008} \cite{Concas:2010}.
Tosun et al. applied Social Networks Aanalysis to OO software metrics source code to assess defect prediction performance of these metrics \cite{Tosun:2009}

Product metrics, extracted by analyzing static code of software, have been used to build models  that relate these metrics to failure-proneness 
\cite{McCabe:76} \cite{Subramanyam:2003} \cite{CK:1998} \cite{Basili:1996} \cite{Gyimoothy:2005}.
Among these, the CK suite is historically the most adopted and validated to analyze bug-proneness of software systems 
\cite{Subramanyam:2003} \cite{CK:1998} \cite{Basili:1996} \cite{Gyimoothy:2005}.
CK suite was adopted by practitioners \cite{CK:1998} and is also incorporated into several industrial software development tools. 
Based on the study of eight medium-sized systems developed by students, Basili et al. \cite{Basili:1996} were among the first to find that OO metrics 
are correlated to defect density.
Considering industry data from software developed in C++ and Java, Subramanyam and Krishnan \cite{Subramanyam:2003}
showed that CK metrics are significantly associated with defects. 
Among others, Gimothy et al. \cite{Gyimoothy:2005}, studying a Open Source system, validated the usefulness of these metrics for fault-proneness prediction.\\ 
CK metrics are intended to measure the degree of coupling and cohesion of classes in OO software contexts.
Statistical analysis has also been used in literature to detect tipical features of complex software and to relate 
the statistical properties to software quality. 

Recently, some researchers have started to study the field of software, in the perspective of finding and studying the associated power-law distributions.
In fact, many software systems have reached such a huge dimension that it looks sensible to treat them using the stochastic
random graph approach~\cite{Focardi:2000}.

Examples of these properties are the lines of code of a class, a function or a method; the number of times a function or a method is called in the system; the number of time a given name is given to a method or a variable, and so on.

Some authors already found significant power-laws in software systems. 
Cai and Yin \cite{IS1:2009} found that the degree distribution of software execution processes 
may follow a power-law or display small-world effects. 
Potanin et al.~\cite{Potanin:2005} showed that the graphs formed by run-time objects, and by the references between them in object-oriented applications, are characterized by a power-law tail in the distribution of node degrees. Valverde et al. \cite{Valverde:2002}\cite{Valverde:2003} found similar properties studying the graph formed by the classes and their relationships in large object-oriented projects. They found that software systems are highly heterogeneous small world networks with scale-free distributions of the connection degree. Wheeldon and Counsell \cite{Wheeldon:2003} identified twelve power laws in object-oriented class relationships of Java programs. In particular, they analyzed the distribution of class references, methods, contructors, field and interfaces in classes, and the distribution of method parameters and return types. Myers \cite{Myers:2003} found analogue results on large C and C++ open source systems, considering the collaborative diagrams of the modules within procedural projects and of the classes within the OO projects. He also computed the correlation between some metrics concerning software size and graph topological measures, revealing that nodes with large output degree tend to evolve more rapidly than nodes with large input degree. Other authors found power-laws studying C/C++ source code files, where graph edges are the files, while the "include" relationships between them are the links~\cite{Gorshenev:2004}, ~\cite{deMoura:2003}. Tamai and Nakatani~\cite{Tamai:2002}, proposed a statistical model to analyze and explain the distributions found for the number of methods per class, and for the lines of code per method, in a large object-oriented system. 

While most of these studies are based on static languages, such like C++ and Java, Marchesi et al.\cite{Marchesi:2004} provide evidence that a similar behavior is displayed also by dynamic languages such as Smalltalk. Concas et al. found power-law and log-normal distributions in some properties of Smalltalk and Java software systems -- the number of times a name is given to a variable or a method, the number of calls to methods with the same name, the number of immediate subclasses of a given class in five large object-oriented software system ~\cite{Concas:2006},~\cite{Concas:2007}. The Pareto principle is used to describe how faults in large software systems are distributed over modules \cite{Fenton:2000}, \cite{Ostrand:2002}, \cite{Ostrand:2005}, \cite{Andersson :2007}, \cite{Zhang:2008}. Baxter et al. \cite{baxter:2008} found power-law and lognormal distributions in the class relationship in Java programs. They proposed a simple generative model that reproduces the features observed in real software graph degree distributions. Ichii et al. \cite{Ichii:2008} investigated software component graphs composed of Java classes finding that in-degree distribution follows the power law distribution and the out-degree distribution does not follow the power-law. Louridas et al. \cite{Louridas:2008}, in a recent work, show that incoming and outgoing links distributions have in common long, fat tails at different levels of abstraction, in diverse systems and languages (C, Java, Perl and Ruby). They report the impact of their findings on several aspects of software engineering: reuse, quality assurance and optimization.

We choose to investigate these properties not only because they show a patent power-law behaviour, 
but also because the former two are related to design and coding guidelines, while the last one
is Chidamber and Kemerer (CK) NOC metrics~\cite{CK:1996}.

Wheeldon and Counsell \cite{Wheeldon:2003}, as well as other researchers, found power-laws
in the distributions of many software properties, such as the number of
fields, methods and constructors of classes, the number of
interfaces implemented by classes, the number of
subclasses of each class, as well as the number of classes referenced as field variables and the number of
classes which contain references to classes as field variables. Thus, there is much evidence
that power-laws are a general feature of software systems.
Concas et al.~\cite{Concas:2006, destefanis2016statistical}
explained the underlying mechanism through a model based on a single Yule process in place 
during the software creation and evolution \cite{Yule}. 
Micro patterns represent design at class level and has been studied in software engineering \cite{concas2013micro, ortu2015could, destefanis2012micro,murgia2014influence} and are promising for Smart contract as well, they could catch design decisions that could be associated to good or bad programming practices.

More recently affect metrics have been investigated revealing how during software development productivity and software quality can be 
highly influenced by developers moods \cite{affect1,affect2,affect3,destefanis2016software, ortu2017diverse}, analyzing in deep details the properties of affect data in software development \cite{destefanis2017randomness, ortu2016emotional,ortu2016arsonists, ortu2015jira,ortu2015mining} and building tool for recognize emotions in software developers' text \cite{ortu2015could,ortu2016diverse,ortu2017connecting,murgia2017exploratory}.

\section{Experimental Set-Up}
Etherscan \cite{etherscan} is a web based platform which allows for Ethereum blockchain exploration of all blockcahin addresses.
It allows to recover Smart Contracts bytecode, ABI, and it collects also Smart Contract source codes in Solidity, 
for more than 12000 contracts at time of writing. 
We parsed blockchain addresses related to Smart Contracts available source code and systematically 
downloaded Solidity contracts code as well as the bytecode and the infos associated to the ABI. 

After collected and locally stored Solidity code, bytecode and ABI infos we built a code parser 
in order to extract the software metrics of our interest for each smart contract.
We also manually explored the code to get insights into the more relevant informations to eventually 
extract from the data and to get a flavour of the main features of the overall dataset. 
This exlploratory analysis allowed us to note how the same conract code is often replicated and 
deployed to different blockcahin addresses or deployed with very little changes. 
This pattern reveals how many contracts are simply experiments or are deployed to the blockcahin
for testing and then modified according to test's results. These usually appear in a series
of neighbour blockchain blocks. 
The dataset has thus a little bias but the overall effect is negligible in our analysis since 
there are very few cases of replicated Solidity code. 

The dataset source code has been then parsed for computing total lines of code in the  
associated to a specific blockchain address, the number of smart contracts inside a 
single address code (the analogous of classes into java files, eg. compilation units), 
blank lines, comment lines, number of static calls to events, number of modifiers, number of functions, number of 
payable functions, cyclomatic complexity as the simplest McCabe definition \cite{McCabe:76},
number of mappings to addresses. We also computed the size of the associated bytecode 
and of the vector of contract's ABIs. 
These are the Application Binary Interfaces, defining the interface definition of any smart contract,
known at compilation time and static. All contracts will have the interface definitions of any contracts 
they call available at compile-time \cite{wikieth}.
This specification does not address contracts whose interface is dynamic or otherwise known only at run-time.

The data set is structured in order to keep track of the specific Smart Contract address 
so that any blockchain address related Smart Contract metrics (SCEM: smart contract external metrics) 
can be fully analyzed in relationship with the software metrics self contained into the Smart Contract Solidity code 
(SCIM: smart contract internal metrics).
For example it is possible to investigate Smart Contracts calls to or from other addresses, 
Intra-Smart Contract calls, gas consumption, cryptocurrencies exchanges and the similar can be 
related to the SCIM or are independent from them, included bytecode metrics or ABI metrics. 
ABI metrics in particular are the Smart Contract interface and reflect the external exposure of the 
Smart Contract towards blockchain calls from other addresses, which can be interaction with 
other Smart Contracts as well. 

It is worth noting that not all the measures related to addresses stay constant but many of them 
depend on the time of analysis and cannot be defined among the SC metrics, and others can simply be 
contract variables, like the amount of ether stored into the contract, the amount of gas, the number 
of owners in a multiowned contract, the contract performance or popularity in terms of calls 
to the contract and so on. 
In such cases much care is needed in order to evaluate a relationship among Smart Contract 
software metrics and other measures blockchain related, not only because they may be time varying, but also because 
other external factors can be in place. For example, the success of a contract could be defined in terms 
of calls to that contract, but if the contract implements an Initial Coin Offer, then 
most likely the contract in itself, tought as software code, has probably little to do with it. 

For each software metric we computed standard statistics like average, median, maxima and minima
values and standard deviation. Furthermore we verified what kind of statistical distribution 
these metrics belong to. This is particularly important when comparing Smart Contract's source 
code with other source code metrics, eg. Java source code, for standard software projects. 
In fact the literature on software metrics demonstrates that there exist statistical distributions
which are typical of specific metrics regardless the programming language used for software development.
 
In particular LOC, coupling metrics, like fan-in and fan-out, and other software metrics are known to display 
a fat tail in their statistical distribution \cite{Louridas:2008}, \cite{Concas:2007} regardless the programming language, 
the platform or the software paradigm adopted for a software project.

Due to the domain specific constraints the Smart Contract software must satisfy to, in particular 
limited size resources, it is not granted that such software metrics does respect the canonical 
statistical distributions found in general purpose software projects. 
It is one of the aims of this reserch to verify and eventually discuss such conjecture.

\section{results}

With start analyzing centrality and dispersion measures for all the computed metrics, like mean, average, 
median, and standard deviation, interquartile range, total variation range. These statistics provide a summary 
of the overall behavior for the metrics values. In particular for asymetric distributions centrality measures 
differs from one another, and in the case of power laws distributions the largest values of the metrics cam 
be order of magnitude larger than central and low values. 

coomenti su power laws

\begin{table}[!ht]
\centering \caption{\textit{Centrality and dispersion statistics computed for all the Smart Contract software metrics.}}
\label{tab:statistics} \vspace{.3cm}
{\scriptsize
\begin{tabular}{|c|c|c||c|c|c|c|c|}\hline 
Metric      & Mean          & Median  & Std & Max & Min & Iqr & 10th - 90th Prc  \\\hline 
Total Lines   & 316.5      & 201    & 326.9       & 4241    & 2 & 265 & 52-720 \\\hline
Blanks   & 55.0         & 32        & 61.4        & 692     & 0 & 49  & 9-138\\\hline
Function   & 25.9          & 18     & 24.2        & 232     & 0 & 20  & 4-58\\\hline
Payable   &  1.2         & 0        & 2           & 15      & 0 & 2   & 0-3\\\hline
Events    & 4.6            & 3      & 4.4         & 42      & 0 & 4   & 0-10\\\hline
Mapping    & 2.5          & 2       & 2.3         & 31      & 0 & 1   & 0-5\\\hline   
Modifier & 2.3            & 1       & 3.2        & 26       & 0 & 3   & 0-7\\\hline
Contract   & 9.2           & 7        & 8.8        & 93     & 1 & 8   & 2-20\\\hline
Address  & 24.8         & 22        & 19.7        & 176     & 0 & 23  & 3-50\\\hline 
Cyclomatic   & 12.7         & 5     & 21.4        & 537     & 1 & 12  & 1-33\\\hline 
Comments   & 77.6       & 42        & 106.5        & 1285   & 0 & 85  & 0-209\\\hline 
ABI     & 3411.4          & 2864    & 2218.5       & 19207  & 0 & 2340 & 917-6332\\\hline 
Bytecode   & 9617.9       & 8273    & 6944.3       & 50607  & 15& 7204 & 2693-18808\\\hline 
LOC   & 183.8          & 121        & 190.2        & 2318   & 2 & 167  & 34-389\\\hline 
\end{tabular}}
\end{table}

Many minima values reult set to zero, since there are a few contracts with almost no code.  
The results on central tendency measures in Tab. \ref{tab:statistics} show that the mean is constantly larger than the median, 
(almost always of about two third) which is a feature typical of right skewed distributions. One simple reason 
explaining this fact is the lower bound posed to all the metrics by the fact that they are defined null or positive, while, 
in principle, large valus are not bounded. A little exception is represented by the Bytecode metric which features values 
for mean and median very close to each other, suggesting a distribution shape which may be not really skewed.
Standard deviations are all comparable with the mean, meaning a large dispersion of values around the last, but there are not 
cases where it is much large than the mean or the media. Values of standard deviation much larger than the mean might be instead 
the case for power laws ditributions and such behavior has already been observed in software metrics for typical software systems
\cite{Wheeldon:2003}, \cite{Concas:2006}.

The maxima are all much larger than the corresponding means and medians, but rarely reach 
one order of magnitude larger and never two orders of magnitude. Finally the 90th percentiles are comparable
with a displacement of some standard deviation from the mean. 
All these results suggest that the selected Smart Contracts metrics might not display a
fat tail or power law distributions which are instead found in the literature for corresponding metrics 
of standard software systems.

\begin{figure}[!ht]
\begin{tabular}{c c}
  \includegraphics[width=0.45\textwidth]{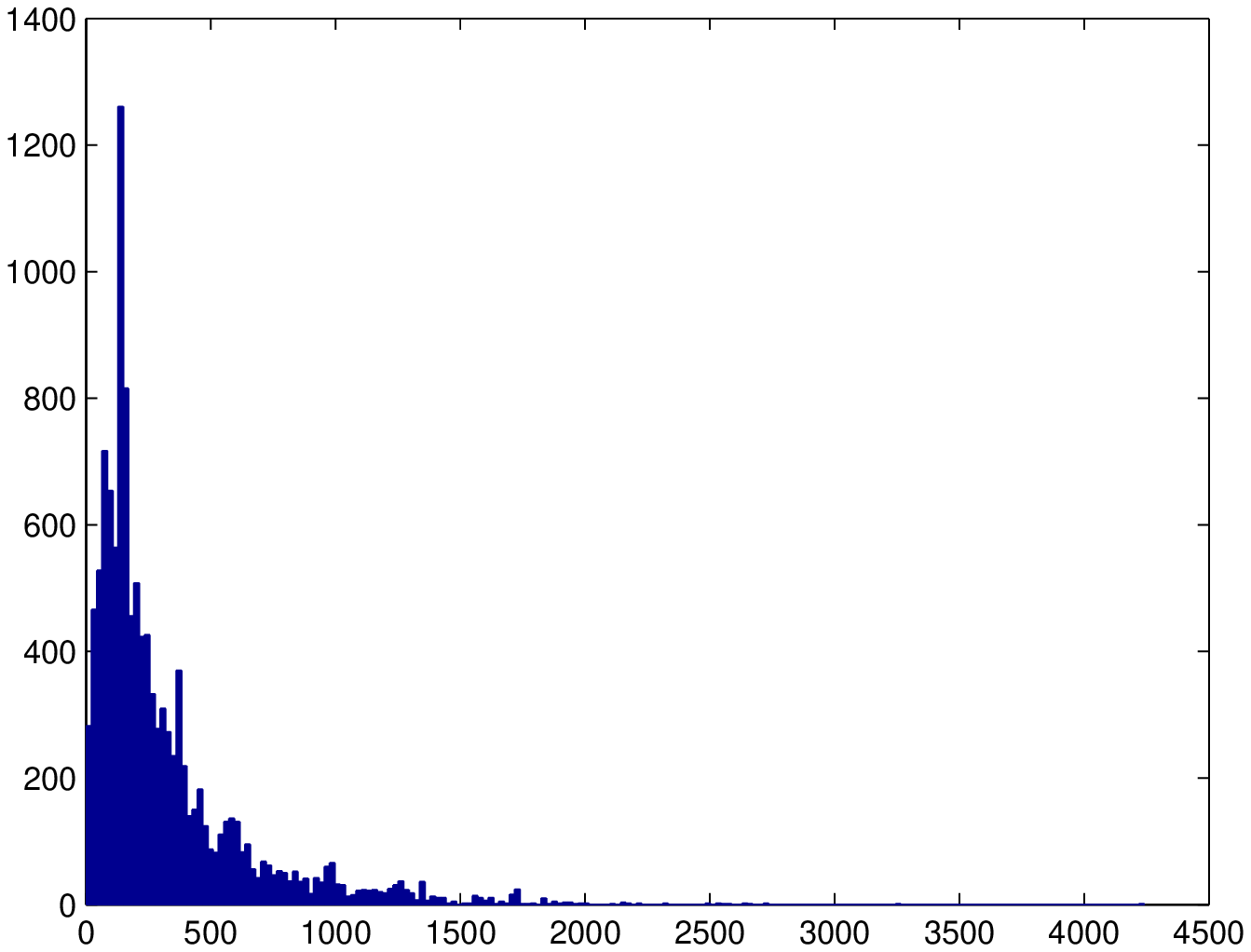}&
  \includegraphics[width=0.45\textwidth]{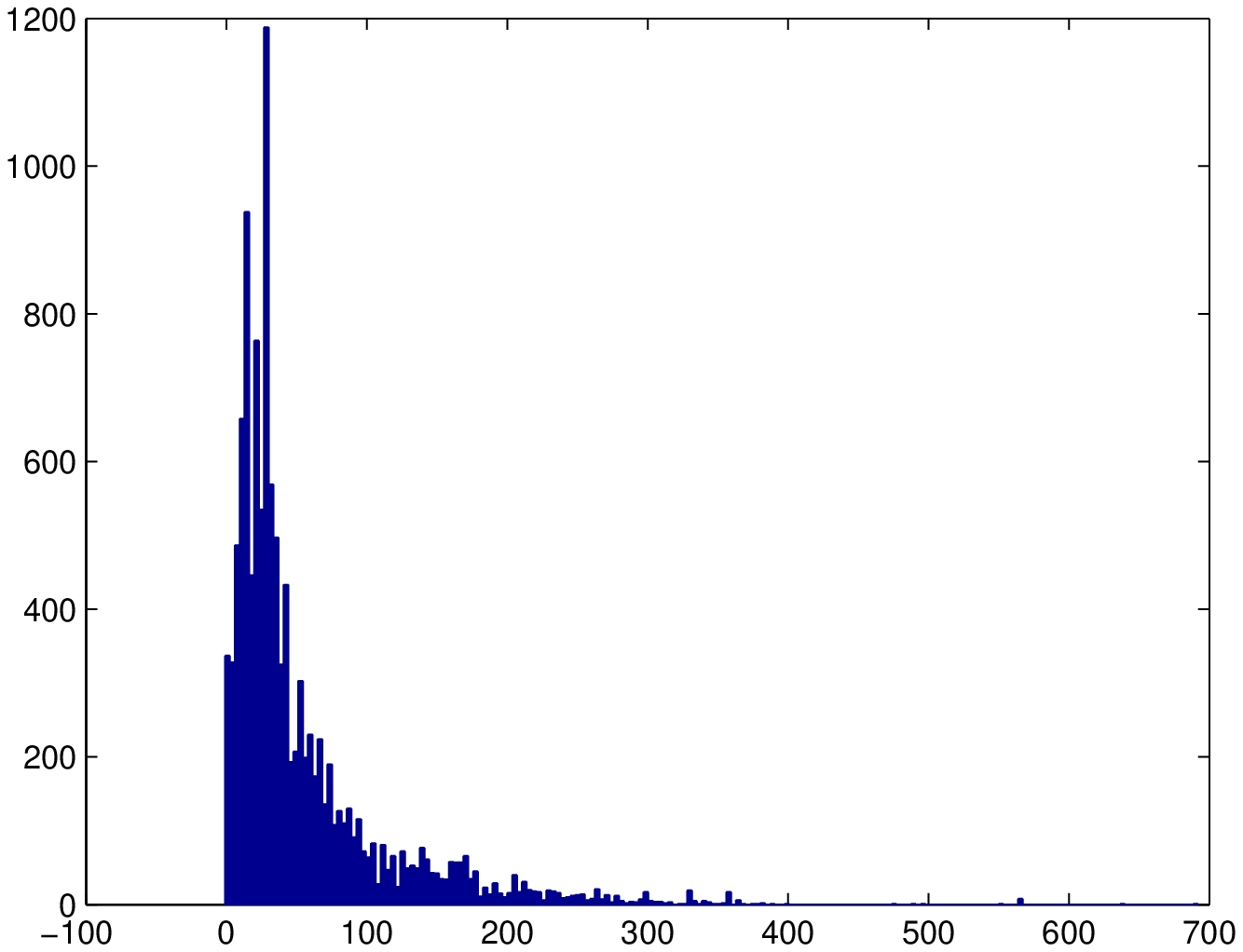} \\
\end{tabular}
\begin{tabular}{c c}
  \includegraphics[width=0.45\textwidth]{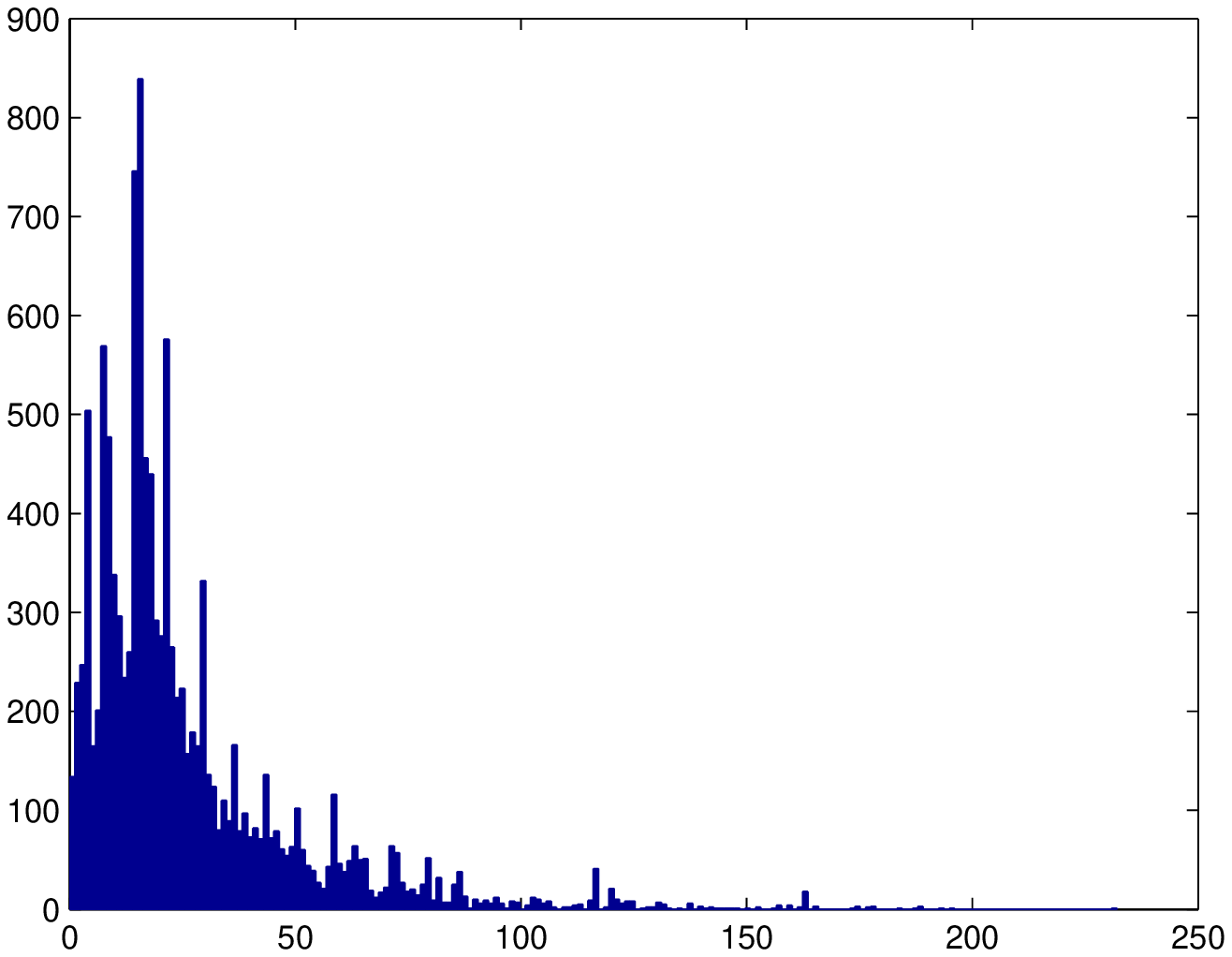}&
  \includegraphics[width=0.45\textwidth]{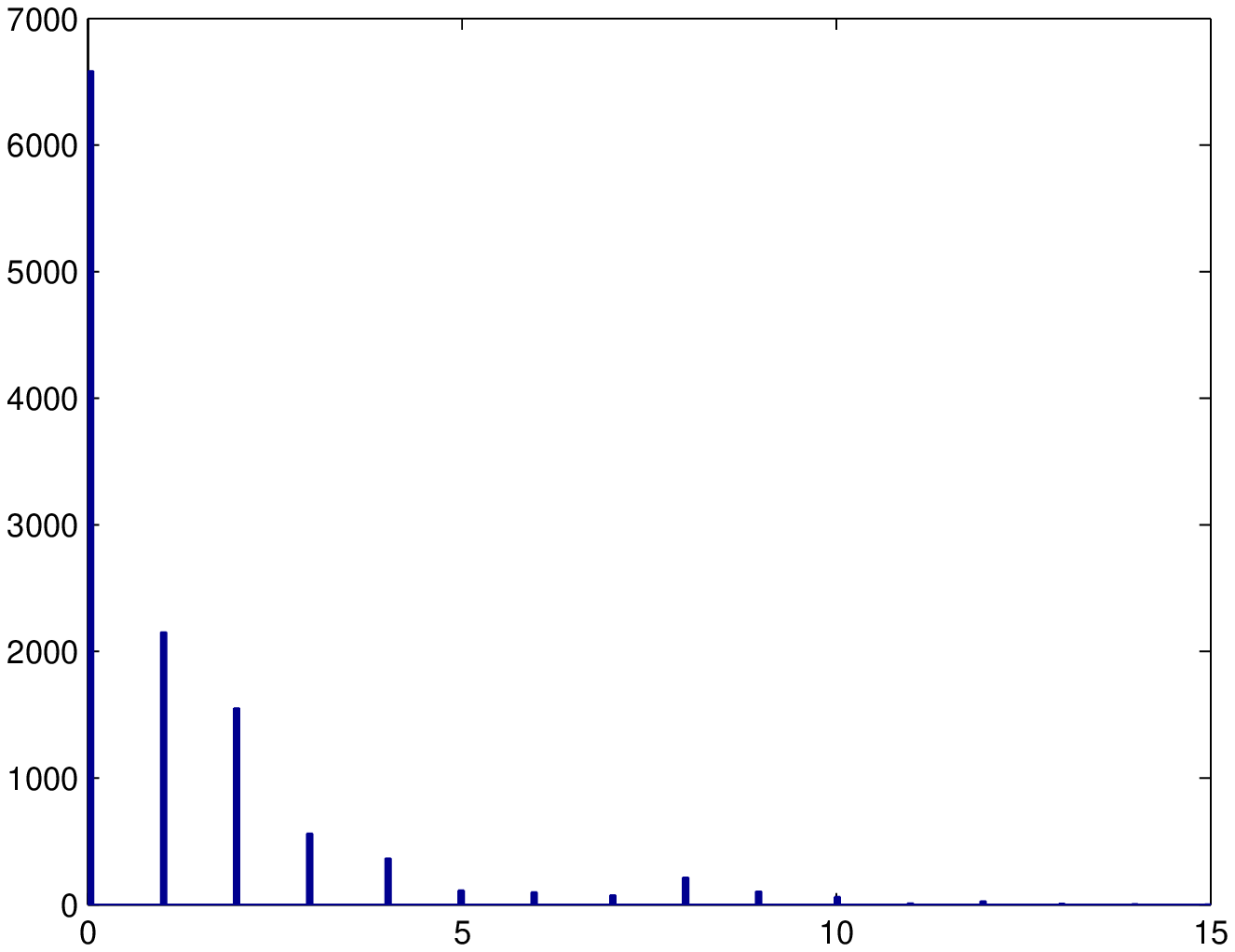}\\
\end{tabular}
\caption{\label{fig:hist1}\textit{Histogram distributions of the metrics Total lines, Blanks, Function and Payable}}
\end{figure}

Nevertheless outlyers values appear for all the metrics and the values in Tab \ref{tab:statistics} are not 
exahustive for explaining completely their statistical properties. 
In order to performe a complete analysis we proceed in two steps. We perform a first qualitative investigation 
analysing the histgrams for all the metrics, then we use more complex statistical models for best 
fitting the Empirical Complementary Cumulative Distribution Function in order to extract quantitative informations
on Smart Contracts software metrics.
The histogram patterns are well known to depend on the bin size and number, as well as on the local density 
of points into the various ranges. Nevertheless they can be an helpful instrument to get insight into the 
distribution shape general features, namely if there may be fat tails, bulk initial distribution values and so on. 
On the contrary the best fittings functions with statistical models provide precise values of core parameters
and can be compared with those reported in literature for standard software metrics.

\begin{figure}[!ht]
\begin{tabular}{c c}
  \includegraphics[width=0.45\textwidth]{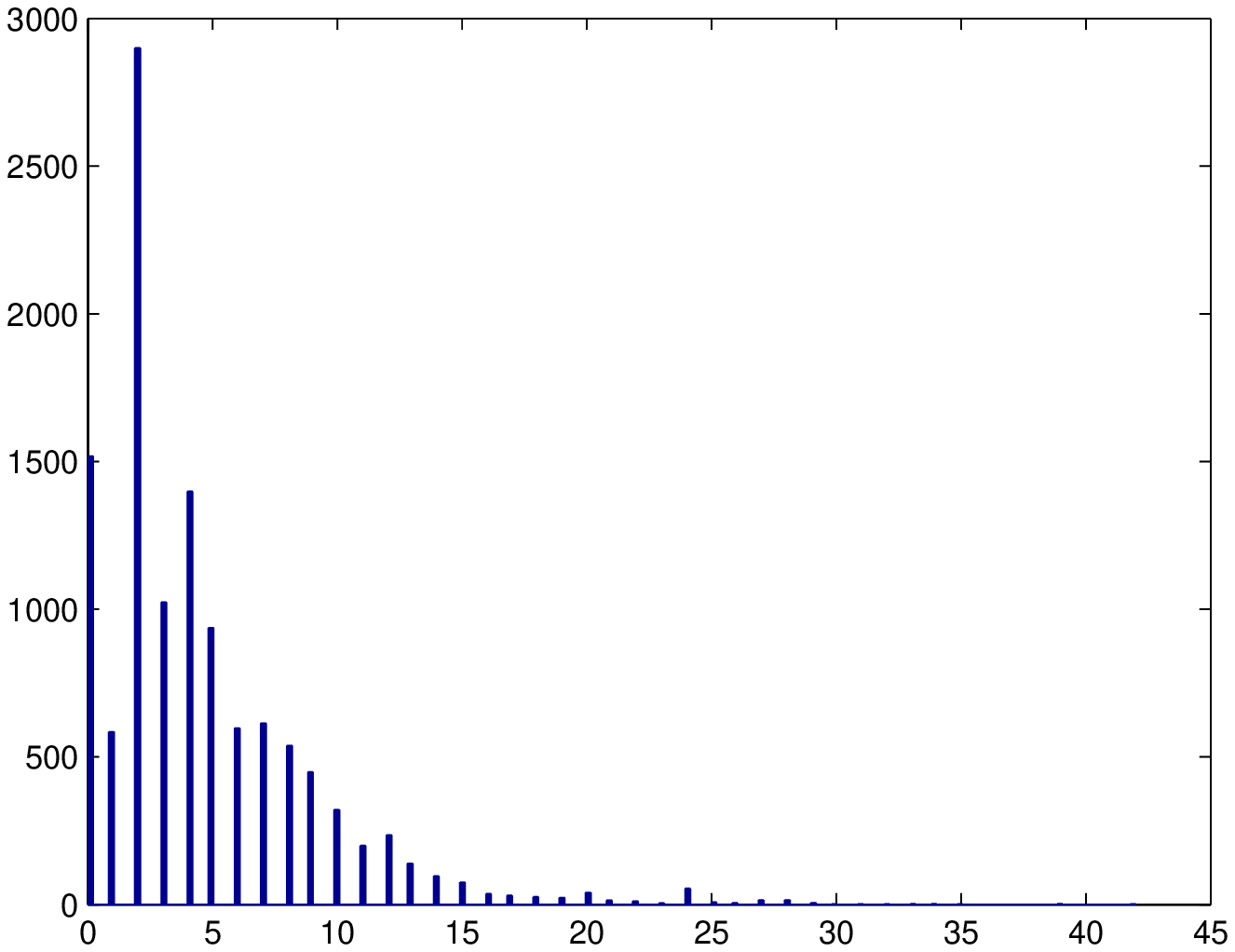}&
  \includegraphics[width=0.45\textwidth]{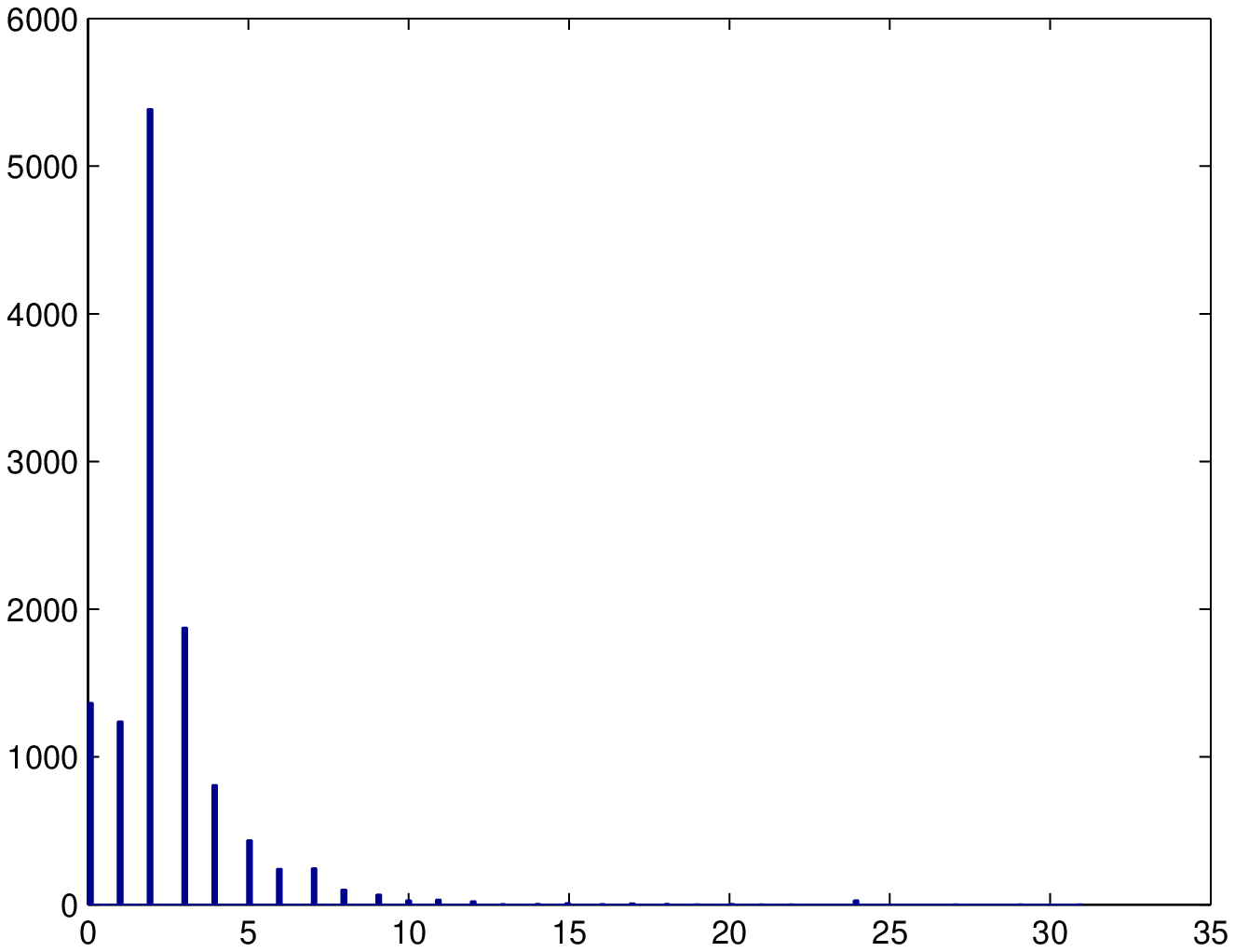}\\
\end{tabular}
\begin{tabular}{c c}
  \includegraphics[width=0.45\textwidth]{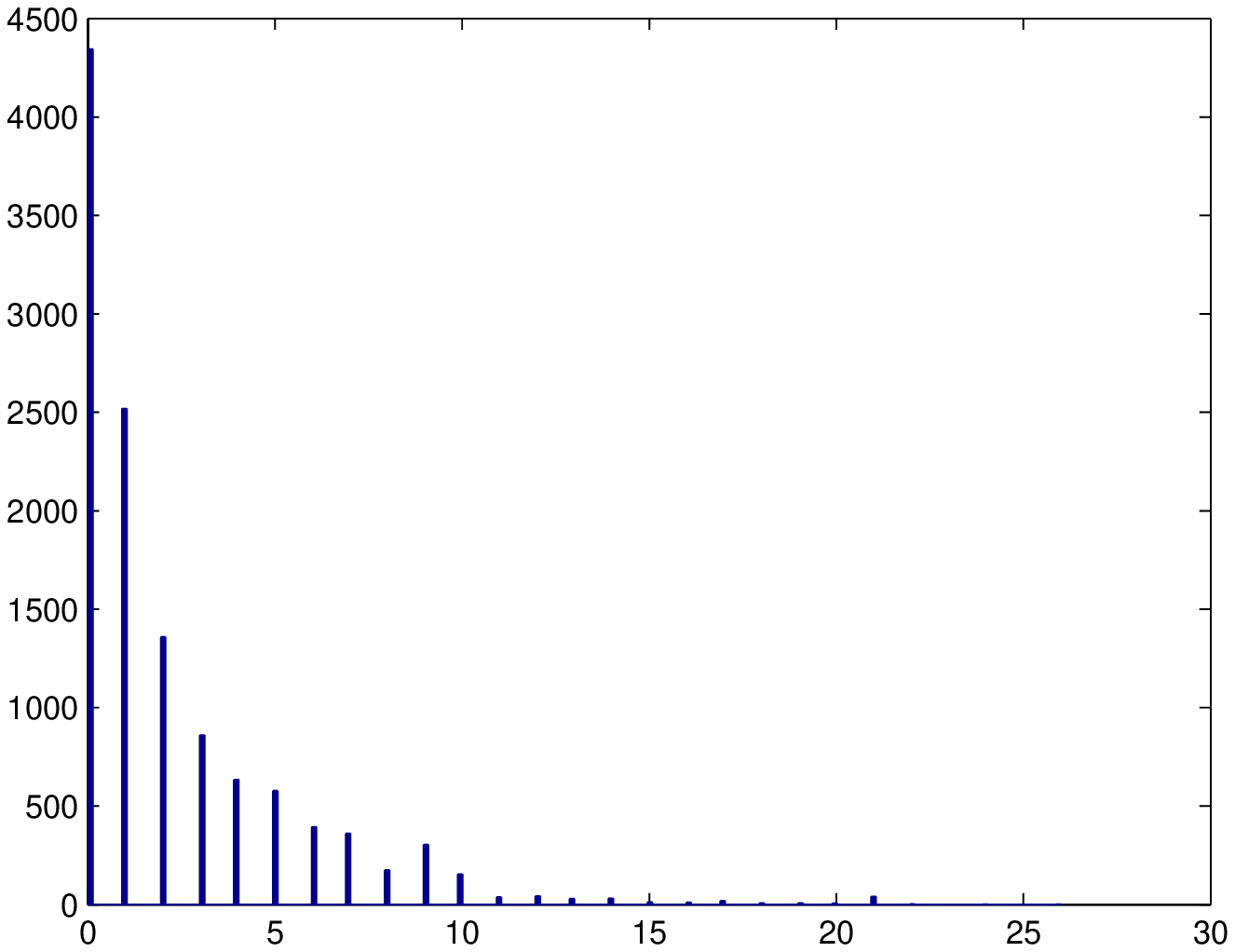}&
  \includegraphics[width=0.45\textwidth]{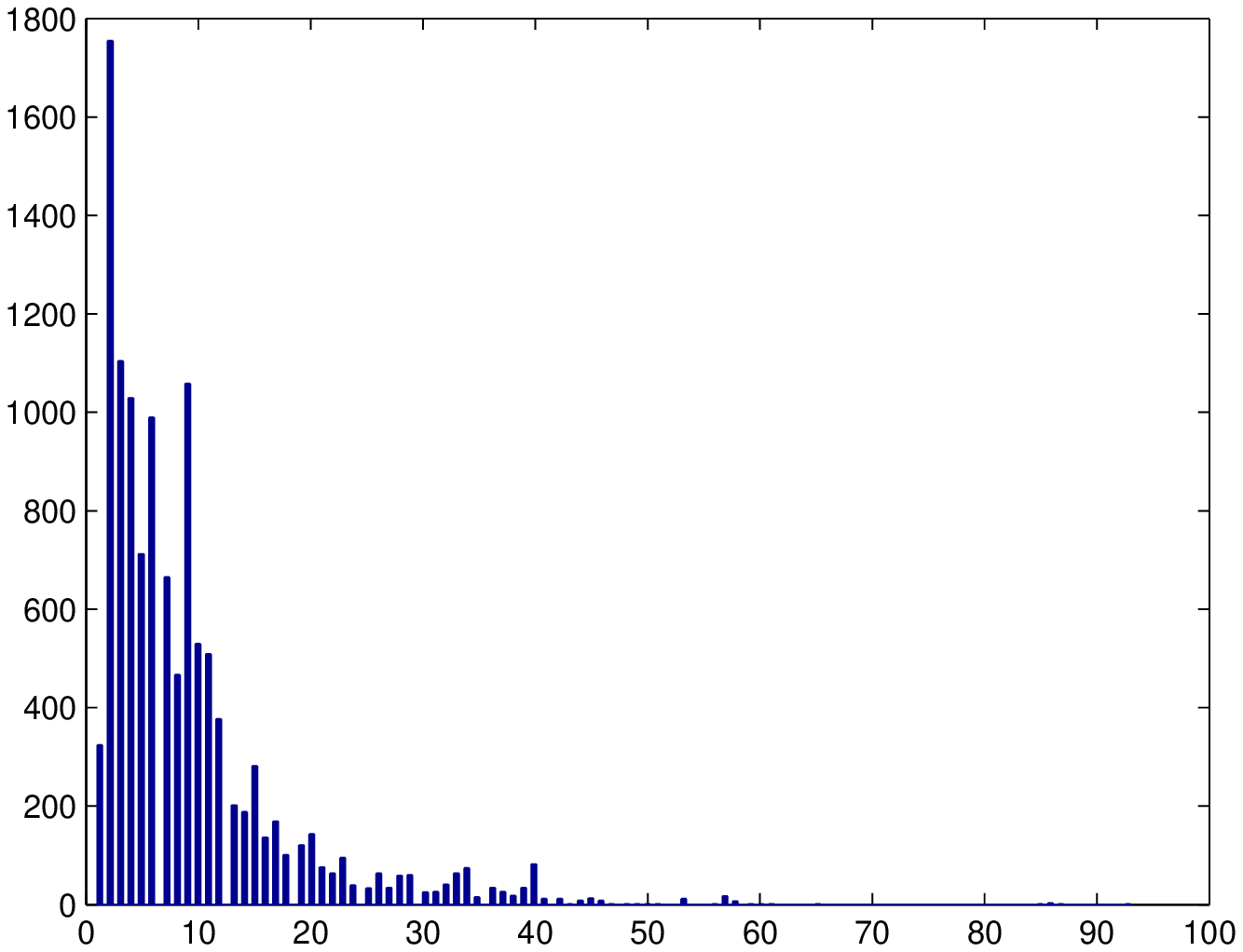}\\
\end{tabular}
\caption{\label{fig:hist2}\textit{
Histogram distributions of the metrics Events, Mapping, Modifier and Contract.}}
\end{figure}

\begin{figure}[!ht]
\begin{tabular}{c c}
  \includegraphics[width=0.45\textwidth]{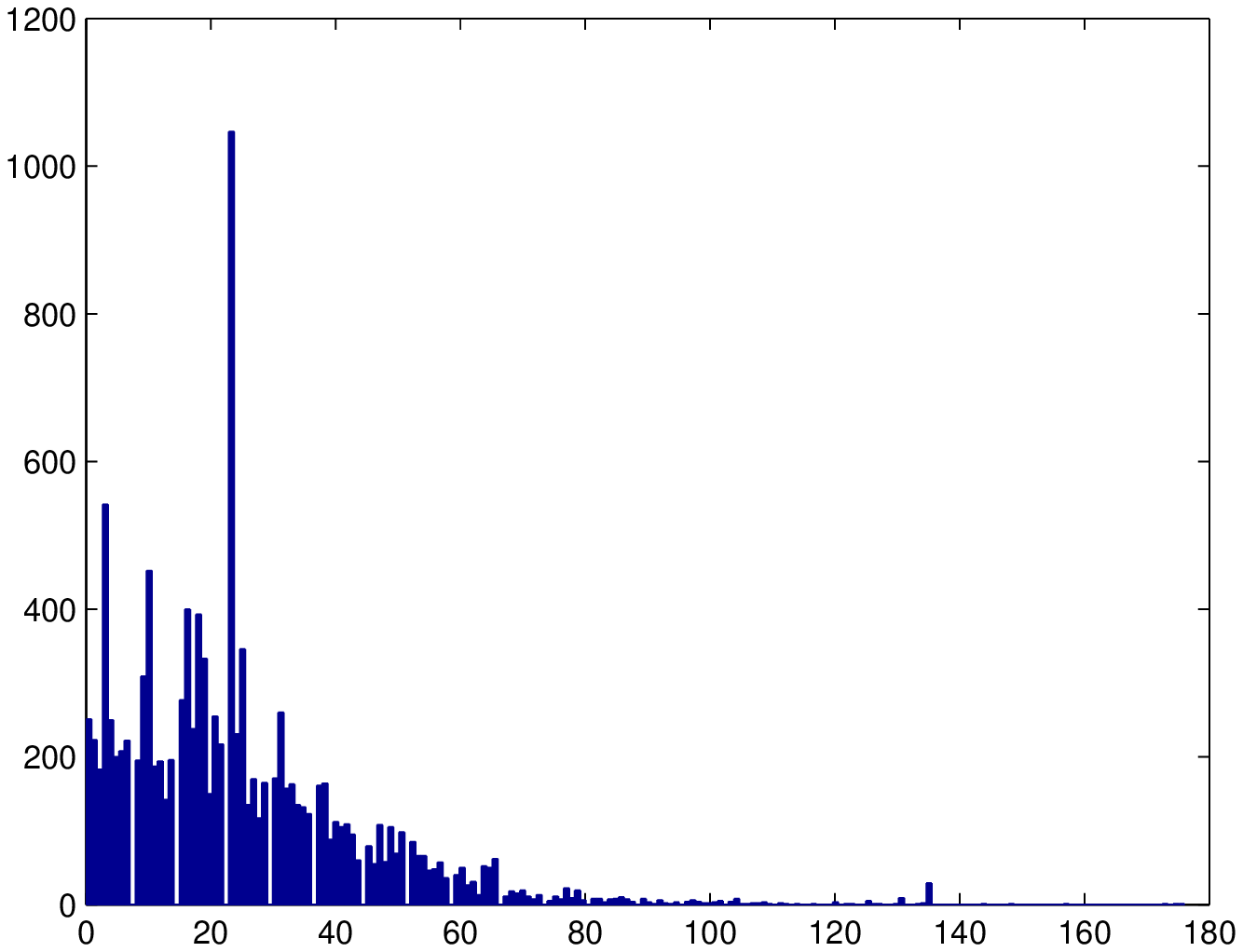}&
  \includegraphics[width=0.45\textwidth]{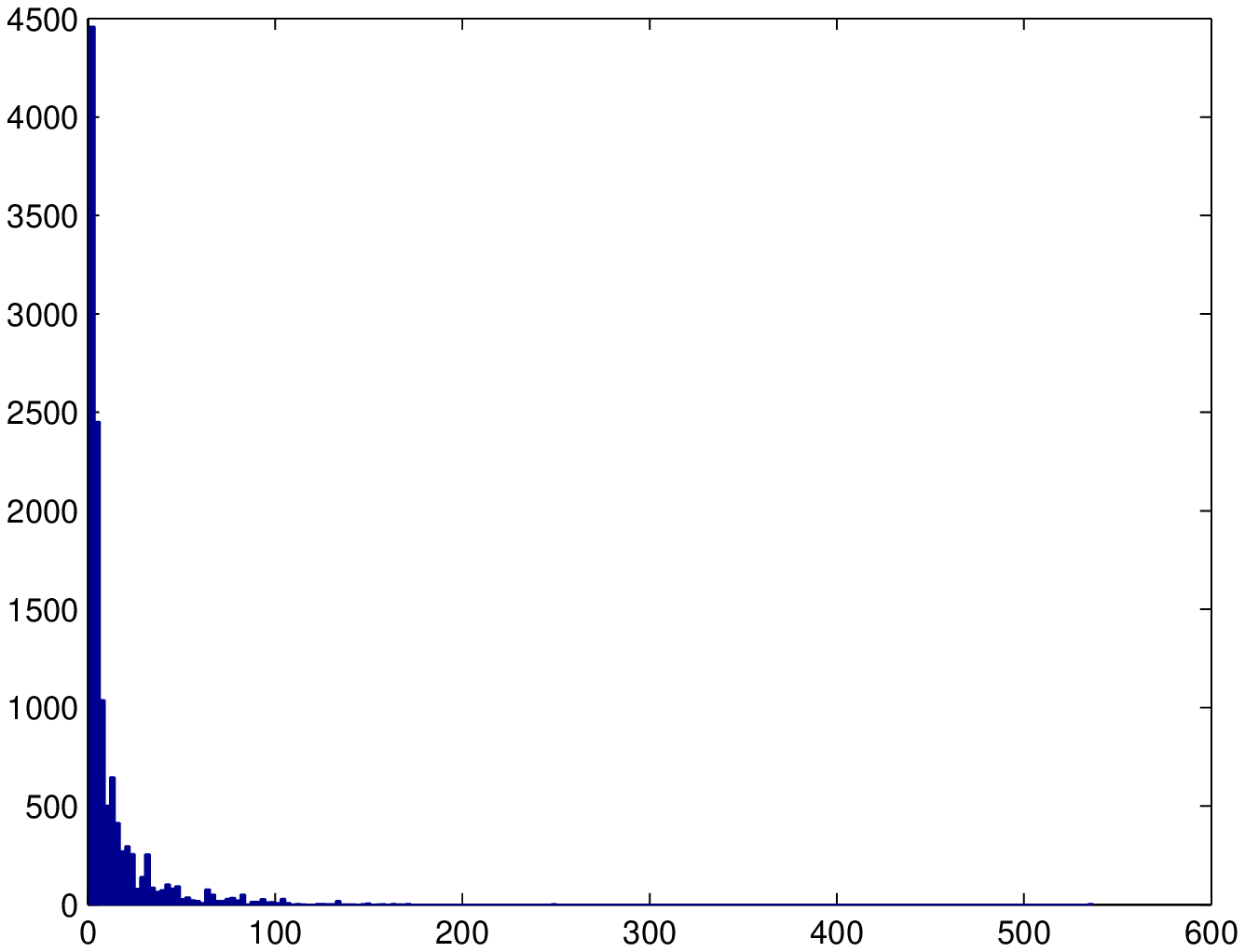}\\
\end{tabular}
\begin{tabular}{c c}
  \includegraphics[width=0.45\textwidth]{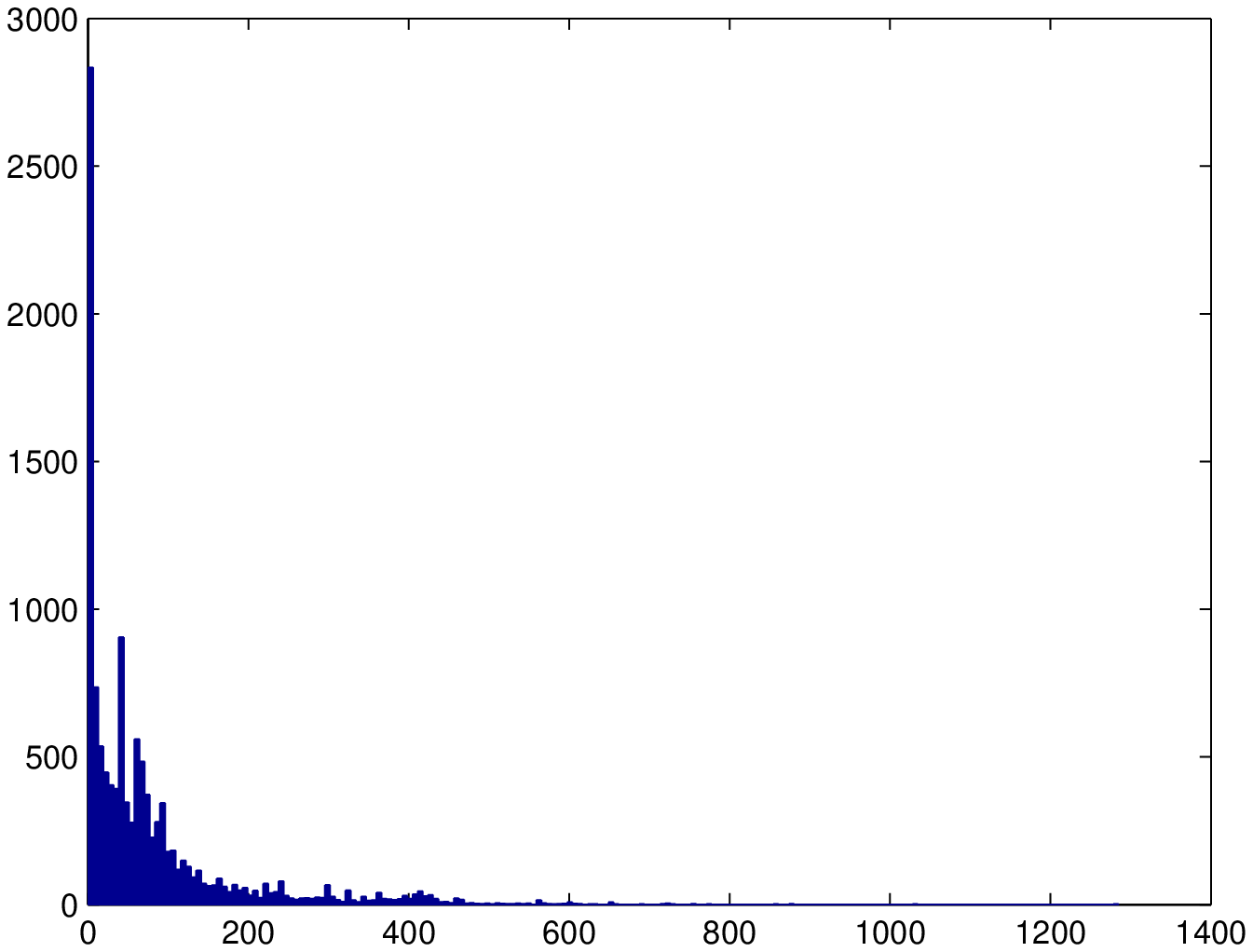}&
  \includegraphics[width=0.45\textwidth]{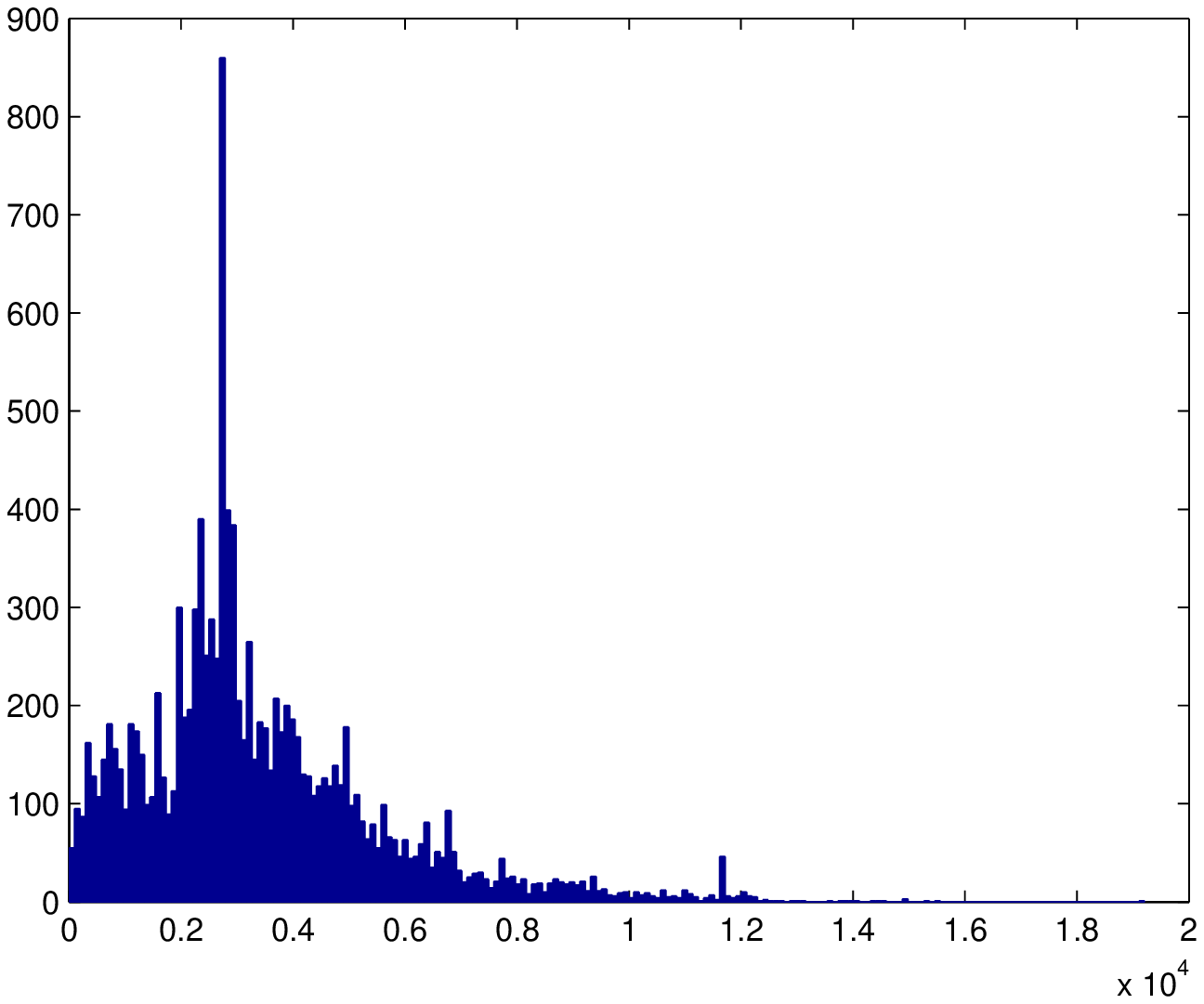}\\
\end{tabular}
\begin{tabular}{c c}
  \includegraphics[width=0.45\textwidth]{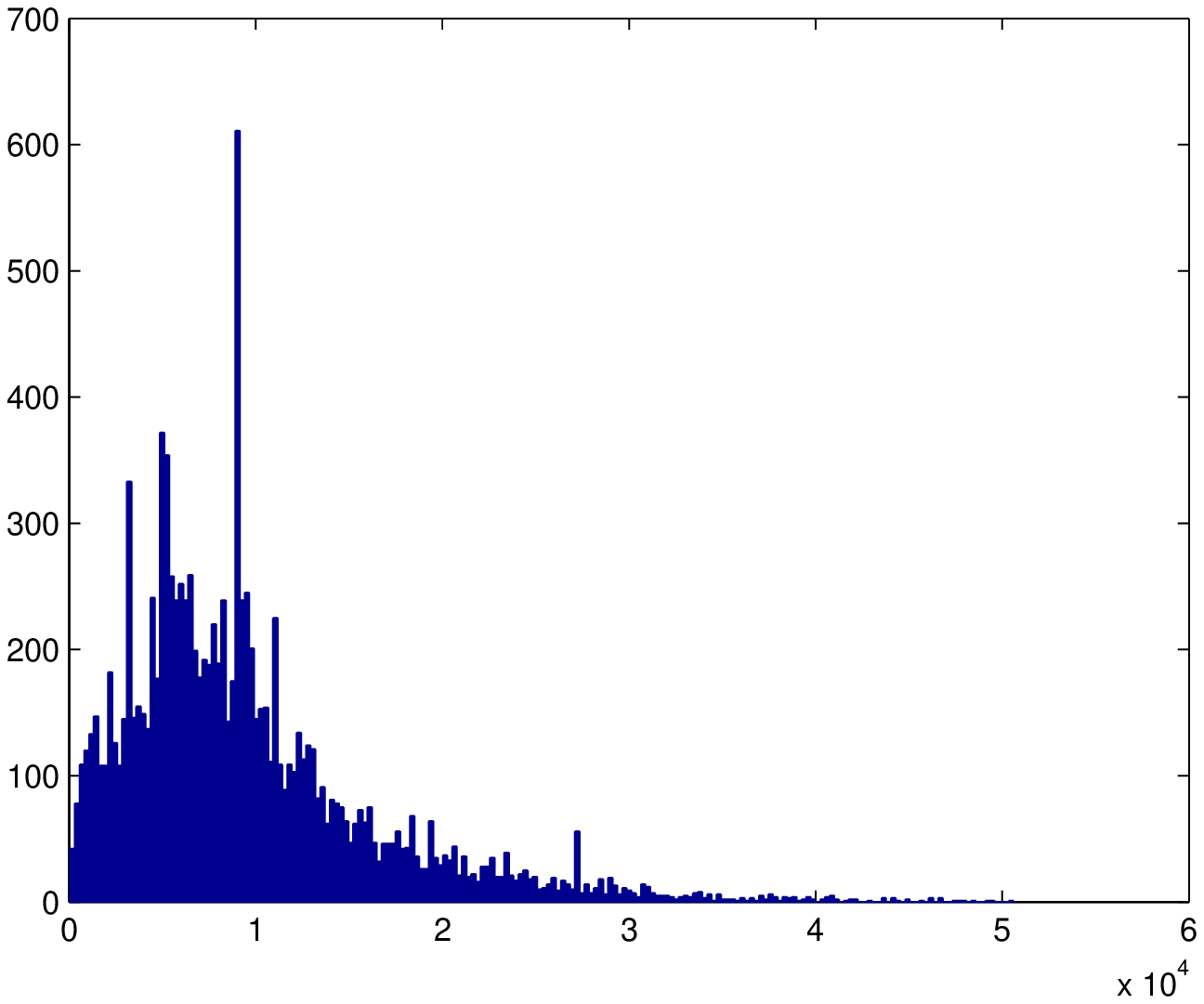}&
  \includegraphics[width=0.45\textwidth]{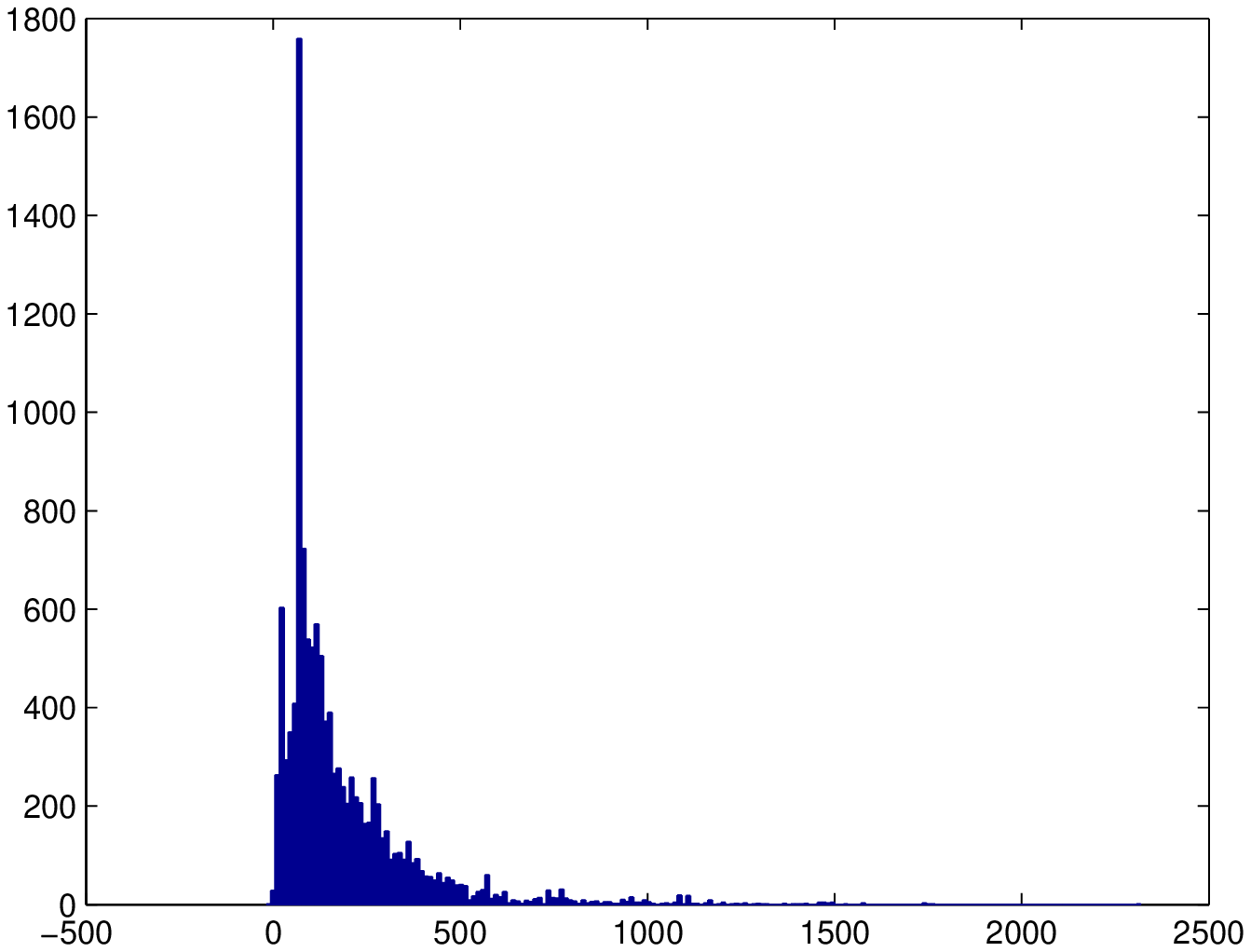}\\
\end{tabular}
\caption{\label{fig:hist3}\textit{Histogram distributions of the metrics Address, Cyclomatic, Comments, ABI, Bytecode and LOCS.}}
\end{figure}

In Figs. \ref{fig:hist1}, \ref{fig:hist2} and \ref{fig:hist3} we report the histograms for all the Smart Contracts software metrics in the 
same order they are reported in Tab. \ref{tab:statistics}. The general shape can be distinguished into two categories. 
From one side there are those metrics whose ranges of variations are quite limited and maximum values are below 50, like Payable, Events, 
Mapping, Modifiable. For such metrics histograms contain too few different values which does not allow to display 
a power law behavior. In particular Payable and Modifiable appear also to have a bell shape which allows to exclude 
a general power law distribution. For Events and Mapping the shape may suggest a power law behavior which is limited 
by the upper bounds reached by the maximum metric values. This deserves to be better investigated using statistical distribution 
modeling.

From the other side the metrics which reach values large enough (whose maxima are over 80) contain enough points
to well populate the histograms. Also in this case many metrics have bell shaped distributions with limited asymmetry and skewness. 
This feature can be ascribed to the limited range of values these metrics can reach. In fact, in cases where 
the metrics can assume virtually arbitrary large values, many orders of magnitude larger that their mean values, 
the bell shape disappear and the shape presents a strong asymmetry with a 
high skewness. This is the behavior observed in literature for metrics in common software systems. 
The only cases where a full power law distribution may approximately hold are those related to the lines of code, like total lines of code, 
blank lines, comments and LOC. But also in these cases the upper bound of the values of the metrics does not allow to fully 
aknoweldge for the power law. This seems to be a structural difference with respect to standard software systems where 
the number of lines of code for a class, for example in Java systems, may easily reach tens of thousands. In fact 
such systems rely on \textit{service classes} containing many methods and code lines, whilst Smart Contracts code 
relies basically on the self contained code.  

It is interesting to note the bell shaped behavior of the ABI metrics and of the Bytecode metric, which strongly 
differ from the shapes associated to lines of code or in general to other metrics. 
In the case of ABI this means that the amount of exposure of Smart Contracts to external interactions has a typical 
scale, provided by clear central values, even if the variance may be quite large. In other words 
Smart Contract exposure to the blockchain is very similar for most of the contracts, with no significative outlyers, 
regardelss the contract size in terms of LOC or other metrics. 
Bytecode display a rather similar but less symmetrics bell shape. In this case the behavior is clearly governed 
by the size contraints imposed by the costs of uploading very large Smart Contracts on the blockchain.

\section{Statistical Modeling}
In order to get insights on the behavior of the statistical distributions underlying Smart Contracts software metrics 
we perform a best fitting analysis using a power law statistical distribution for best fitting the tails 
of the empirical distributions. Furthermore we performed a second analysis making use of the Lognormal statistical model. 
In fact, even when the power law model well represent the data in the tail it usually is unable to best fit the complete 
range of values in the statistical distributions. 

To show the results of such analysis we don't use histograms anymore, which are 
a rough approximation of a Probability Density Function (PDF). 
The histogram representation in fact carries many drowbacks, in particular 
when data are power-law distributed in the tail. 
The problems with representing the empirical PDF are that it is sensitive 
to the binning of the histogram used to calculate the frequencies 
of occurrence, and that bins with very few elements 
are very sensitive to statistical noise. 
This causes a noisy spread of the points in the tail of the distribution, 
where the most interesting data lie. Furthermore, because of the binning, 
the information relative to each single data is lost. 
All these aspects make difficult to verify the power-law behavior 
in the tail.
To overcome these problems from now on we systematically report the experimental CCDF 
(Complementary Cumulative Distribution Function) in log-log scale, as well as the best-fitting curves in many cases. 
This is convenient because, if the PDF (probability distribution function) has a power-law in the tail, the log-log plot displays a straight line for the 
raw data. This is a necessary but by no means a sufficient condition 
for power-law behavior. 
Thus we used log-log plots only for convenience of graphical representation, 
but all our calculations (CDF, CCDF, best fit procedures and the same 
analytical distribution functions we use) are always in normal scale. 
With this representation, there is no dependence on the binning, nor artificial statistical noise  
added to the tail of the data. If the PDF exhibits a power-law, so does
the CCDF, with an exponent increased by one. Fitting the 
tail of the CCDF, or even the entire distribution, results in a 
major improvement in the quality of fit.
An exhaustive discussion of all these issues may be found in \cite{Newman:2005}.
This approach has already been proposed in literature to explain 
the power-law in the tail of various software properties \cite{Concas:2007} 
\cite{Louridas:2008}, \cite{LesHatton}. 

The CCDF is defined as $1 - CDF$, where the CDF (Cumulative Distribution Function) is the integral 
of the PDF. Denoting by $p(x)$ the probability distribution 
function, by $P(x)$ the CDF, and by $G(x)$ the CCDF, we have:  

\begin{eqnarray}
G(x) = 1 - P(x) \\
P(x) = p(X \leq x) = \int_{-\infty}^x p(x')dx' \\
G(x) = p(X \geq x) = \int_x^{\infty} p(x')dx'
\end{eqnarray}

The first distribution we describe is the well-known log-normal distribution.
If we model a stochastic process in which new elements 
are introduced into the system units 
in amounts proportional to the actual 
number of the elements they contain, then the resulting element distribution 
is log-normal. All the units should have the same constant chance for being selected 
for the introduction of new elements \cite{Newman:2005}.
This general scheme has been demonstrated to suit large software systems 
where, during software development, new classes are introduced 
into the system, and new dependencies --links-- among them 
are created \cite{Louridas:2008},  \cite{TSE}. 
The log-normal has also been used to analyze the distribution 
of Lines of Code \cite{Zhang2}. The log-normal distribution has been also proposed in literature 
to explain different software properties (\cite{Noble:2006}, \cite{Concas:2006}, \cite{Louridas:2008}).
Mathematically it is expressed by:

\begin{equation}
p(x) = \frac{1}{\sqrt{2 \pi \sigma} x} e^{ - \left(\frac{ln(x)-\mu}{2 \sigma} \right)^2}
\end{equation}

It exhibits a quasi-power-law behavior for a range of values, and 
provides high quality fits for data with power-law distribution 
with a final cut-off. Since in real data largest values are always 
limited and cannot actually tend to infinity, the log-normal is a very good candidate 
for fitting power-laws distributed data with a finite-size 
effect. Furthermore, it does not diverge for small 
values of the variable, and thus may also fit well the bulk 
of the distribution in the small values range.

The power-law is mathematically formulated as:
\begin{equation}
p(x) \simeq x^{-\alpha}
\end{equation}
where $\alpha$ is the power-law exponent, the only parameter which 
characterizes the distribution, besides a normalization factor. 
Since for $\alpha \ge 1$ the function diverges in the origin, 
it cannot represent real data for its entire range of values. 
A lower cut-off, generally indicated $x_0$, has to be introduced, 
and the power-law holds above $x_0$. Thus, when fitting real data, 
this cut-off acts as a second parameter to be adjusted 
for best fitting purposes. Consequently, the data distribution is said 
to have a power-law in the tail, namely above $x_0$.

\begin{figure}[!ht]
\begin{tabular}{c c}
  \includegraphics[width=0.45\textwidth]{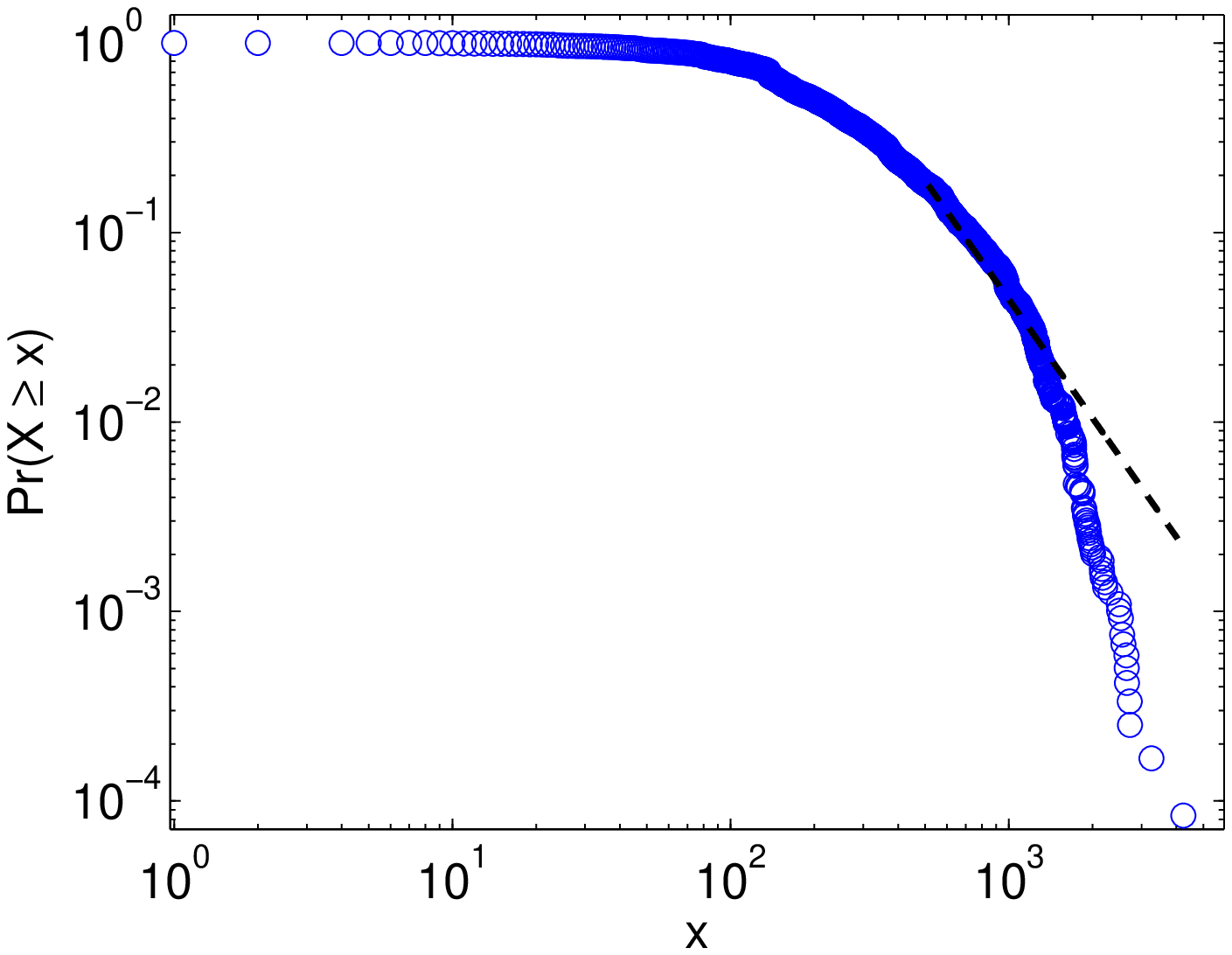}&
  \includegraphics[width=0.45\textwidth]{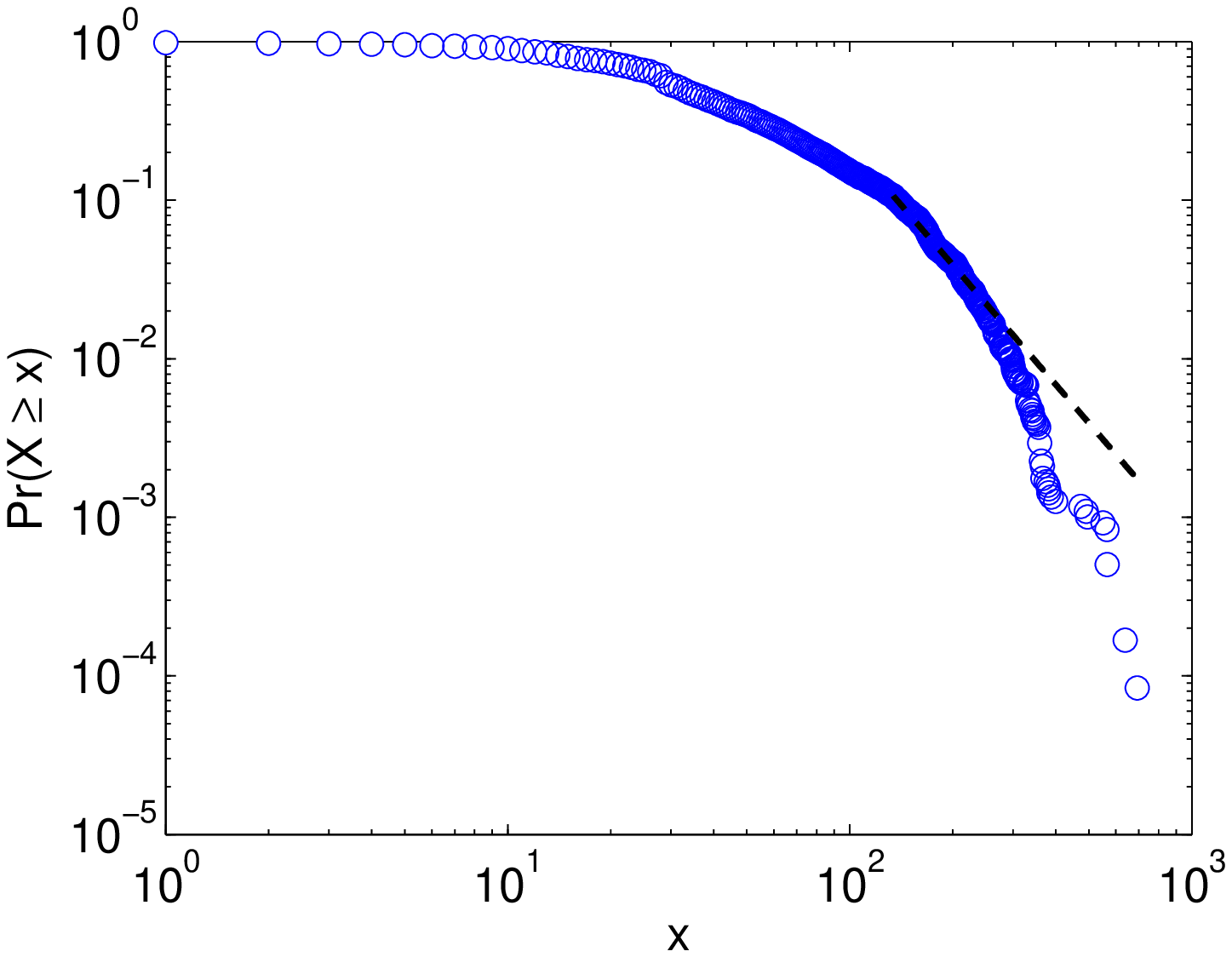} \\
\end{tabular}
\begin{tabular}{c c}
  \includegraphics[width=0.45\textwidth]{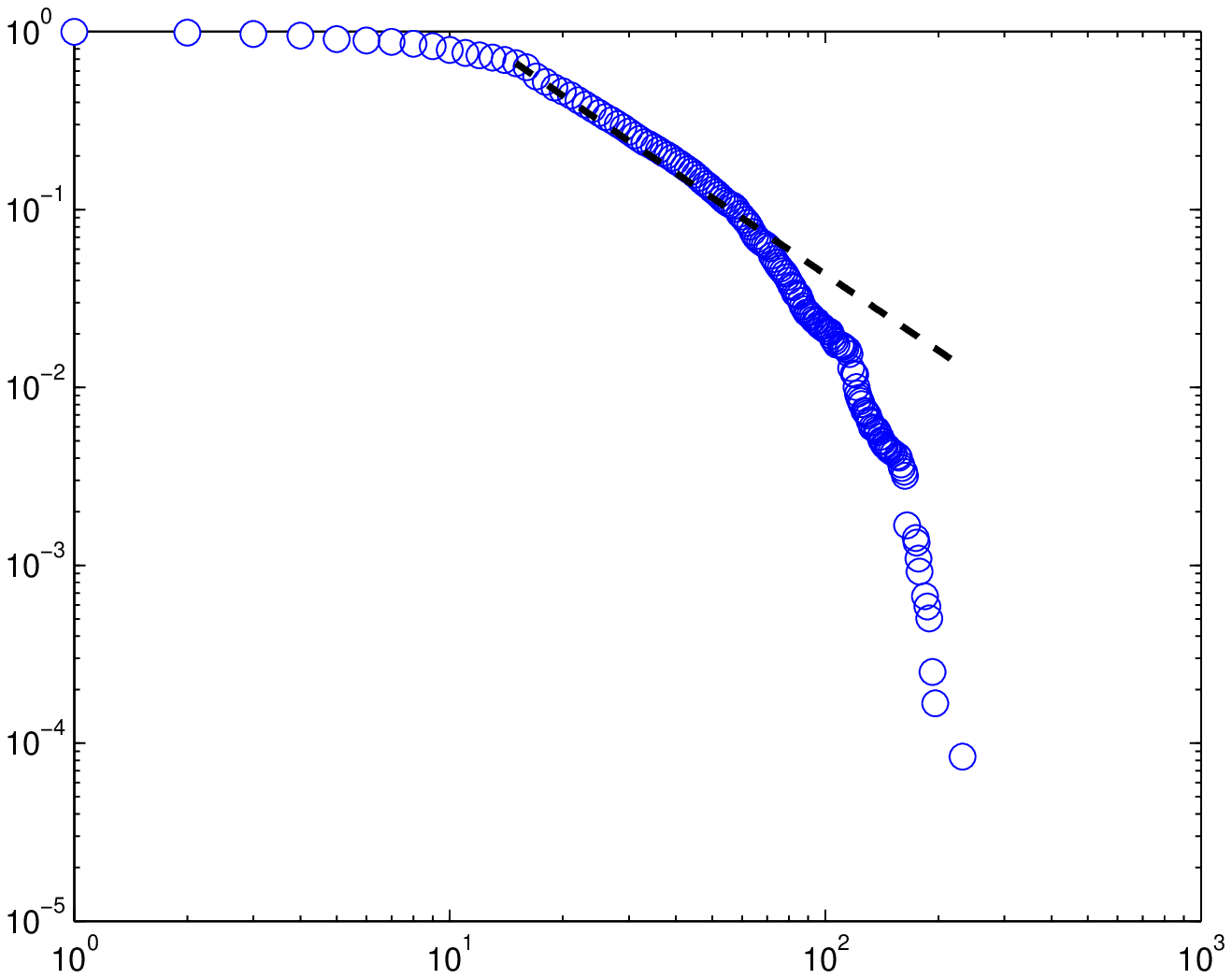}&
  \includegraphics[width=0.45\textwidth]{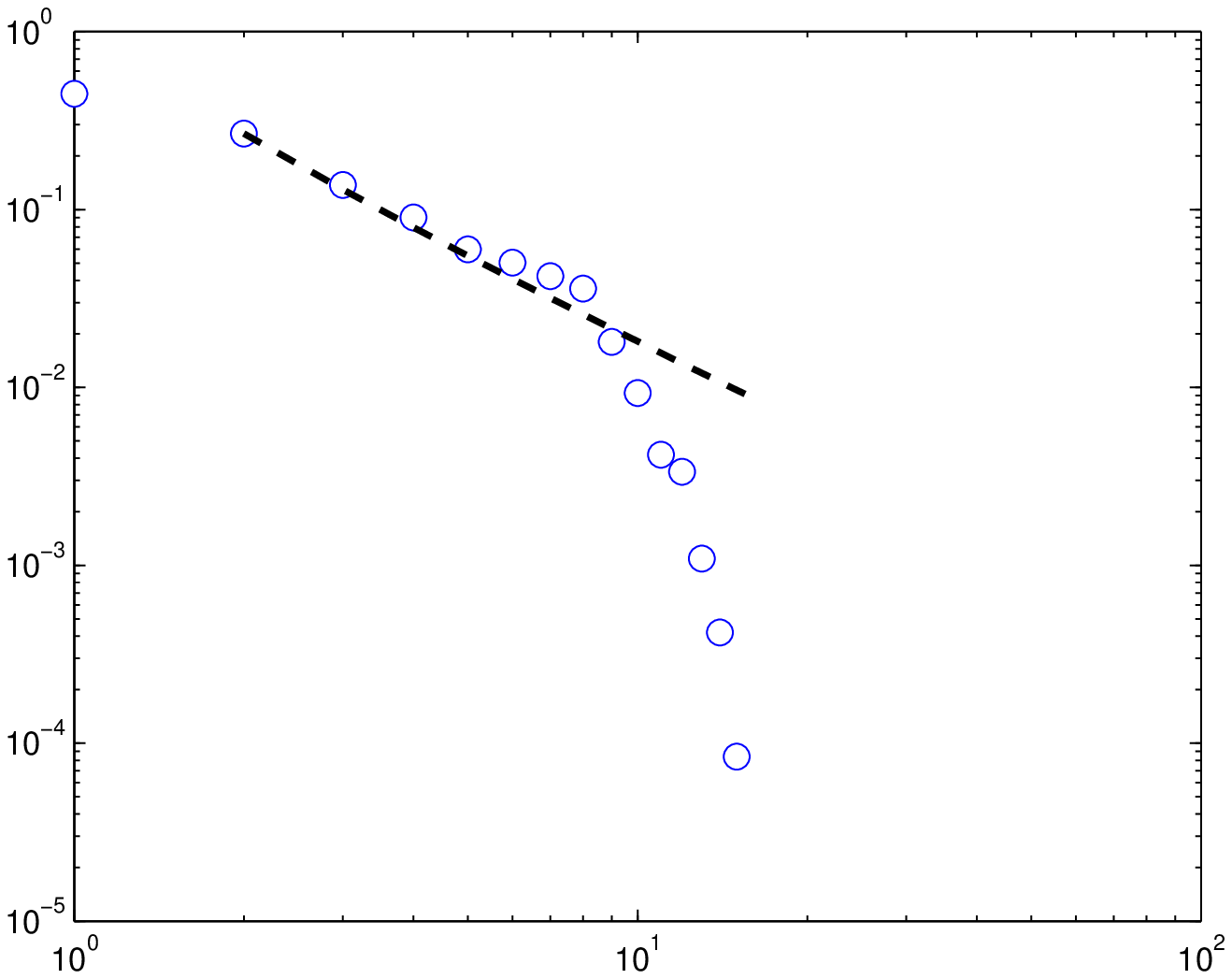}\\
\end{tabular}
\caption{\label{fig:pl1}\textit{Power law best fitting of the metrics Total lines, Blanks, Function and Payable}}
\end{figure}

\begin{figure}[!ht]
\begin{tabular}{c c}
  \includegraphics[width=0.45\textwidth]{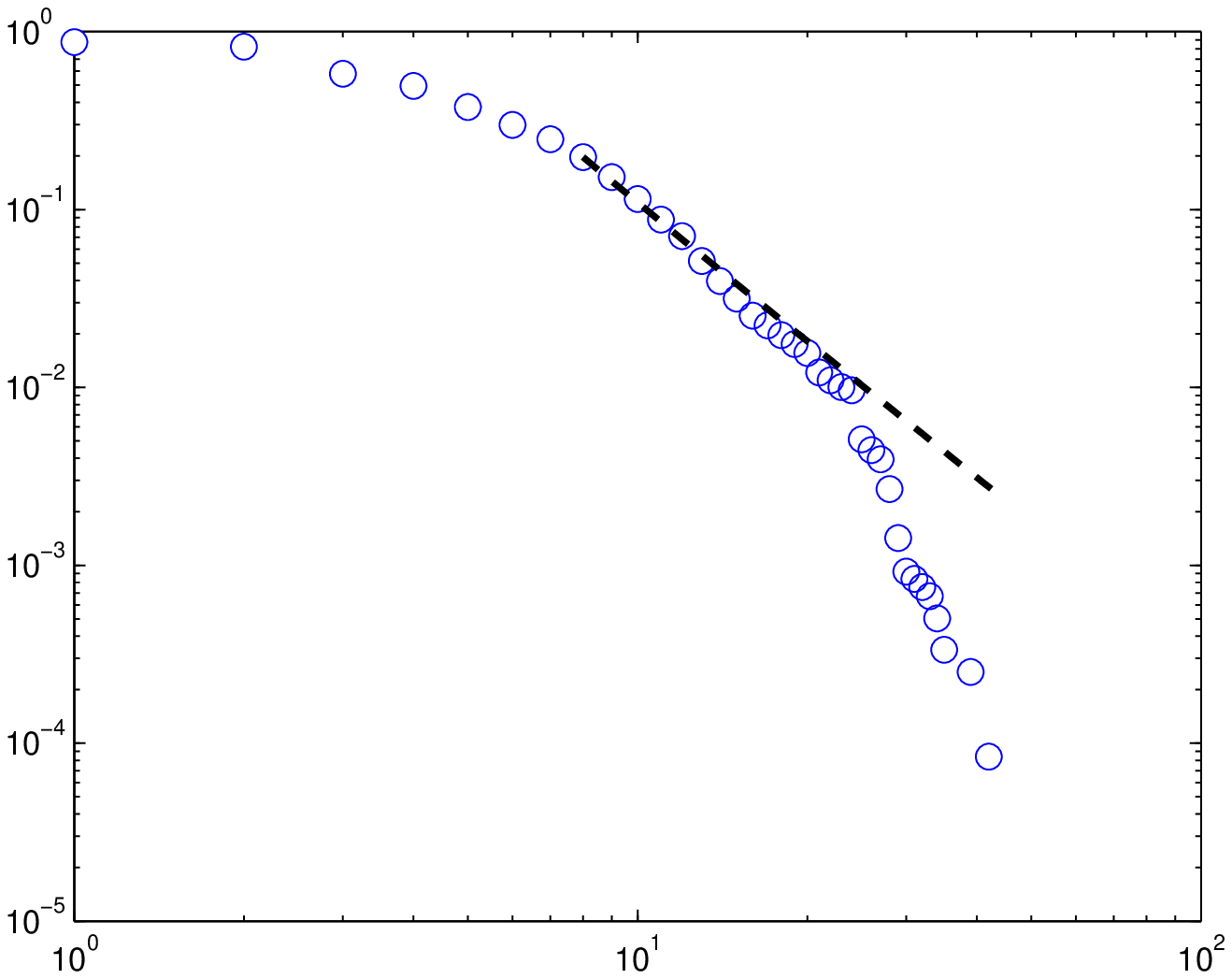}&
  \includegraphics[width=0.45\textwidth]{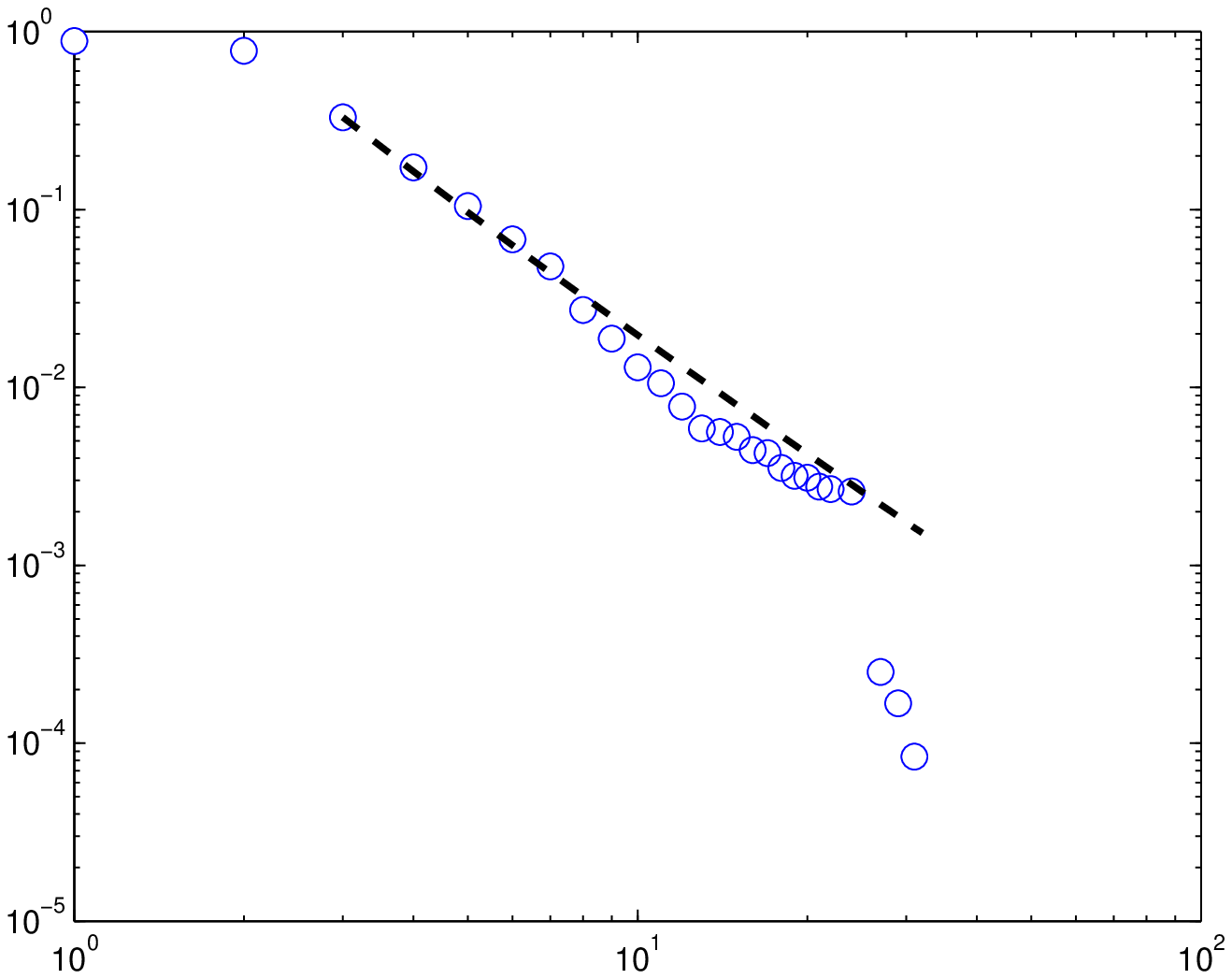} \\
\end{tabular}
\begin{tabular}{c c}
  \includegraphics[width=0.45\textwidth]{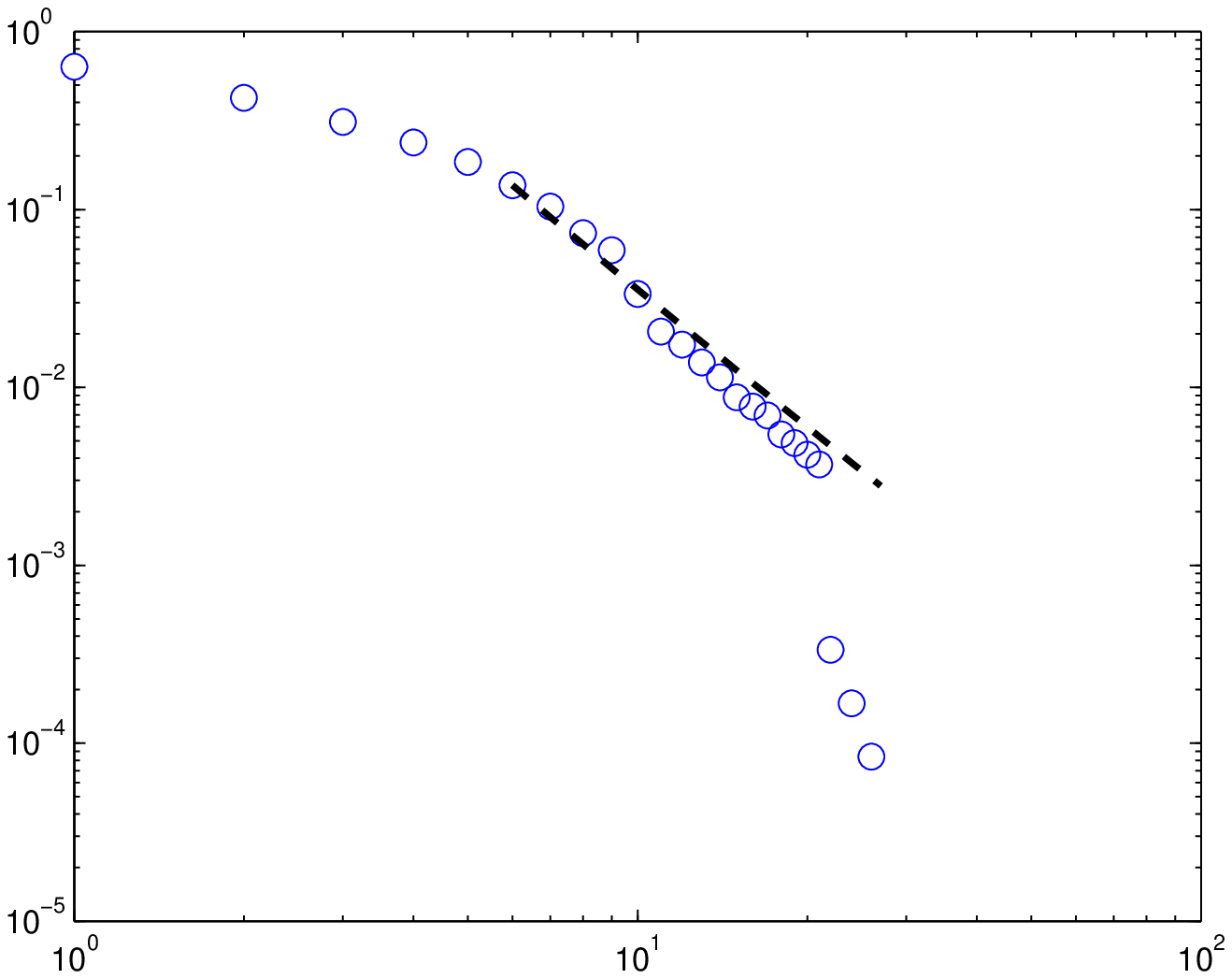}&
  \includegraphics[width=0.45\textwidth]{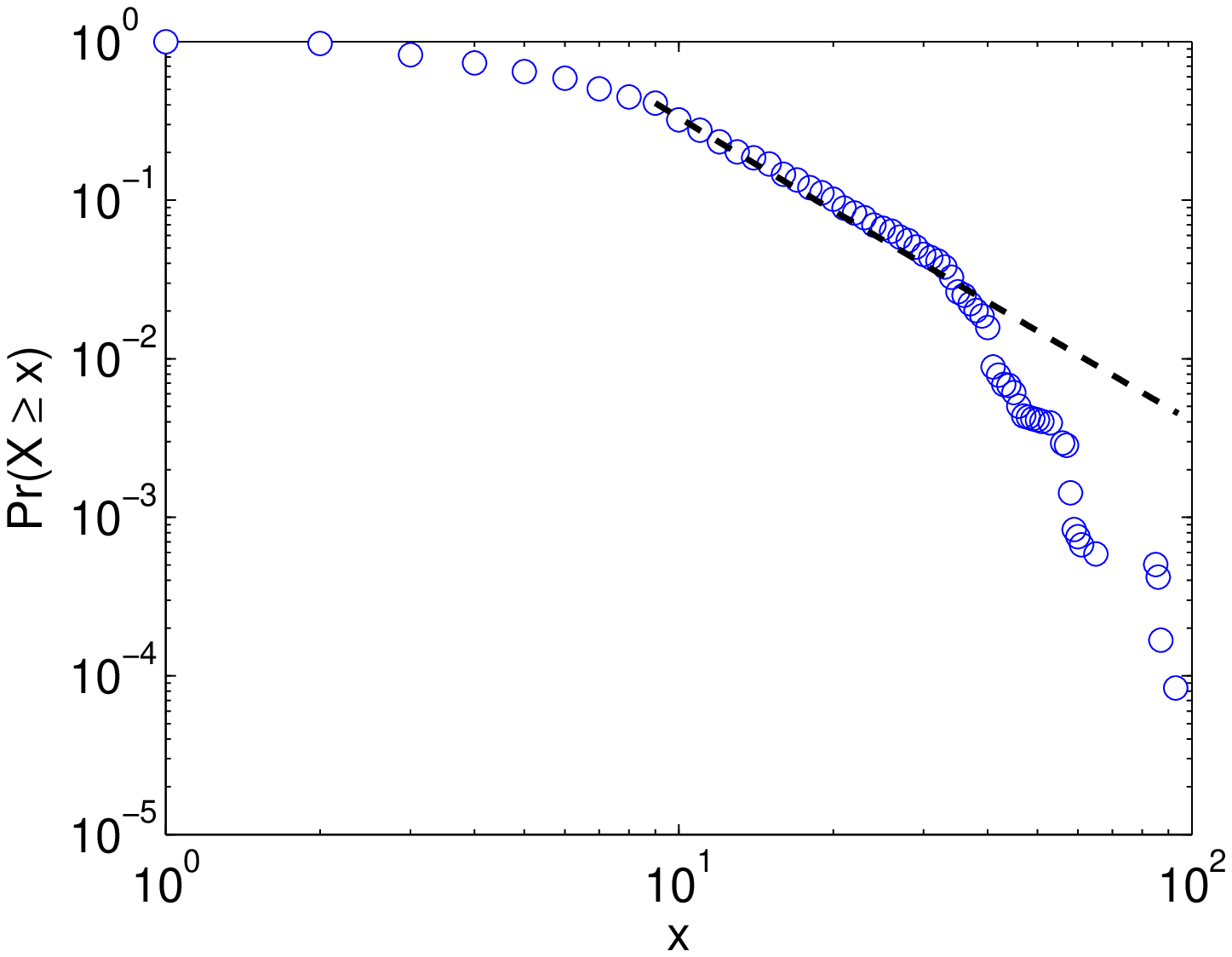}\\
\end{tabular}
\caption{\label{fig:pl2}\textit{Power law best fitting of the metrics Events, Mapping, Modifier and Contract.}}
\end{figure}

In Fig. \ref{fig:pl1} we show the best fitting plot for the power law model for the metrics Total lines, Blanks, Function and Payable. 
The power law in the tail is clearly failed by all metrics.  
In Fig. \ref{fig:pl2} Mapping and Modifier seems to follow a power law, confirmed also by the low values (D $\leq$ 0.05) of the 
Kolmogorof-Smirnov significance test value, but the range where the metrics behave according to a power law regime is too small.

\begin{figure}[!ht]
\begin{tabular}{c c}
  \includegraphics[width=0.45\textwidth]{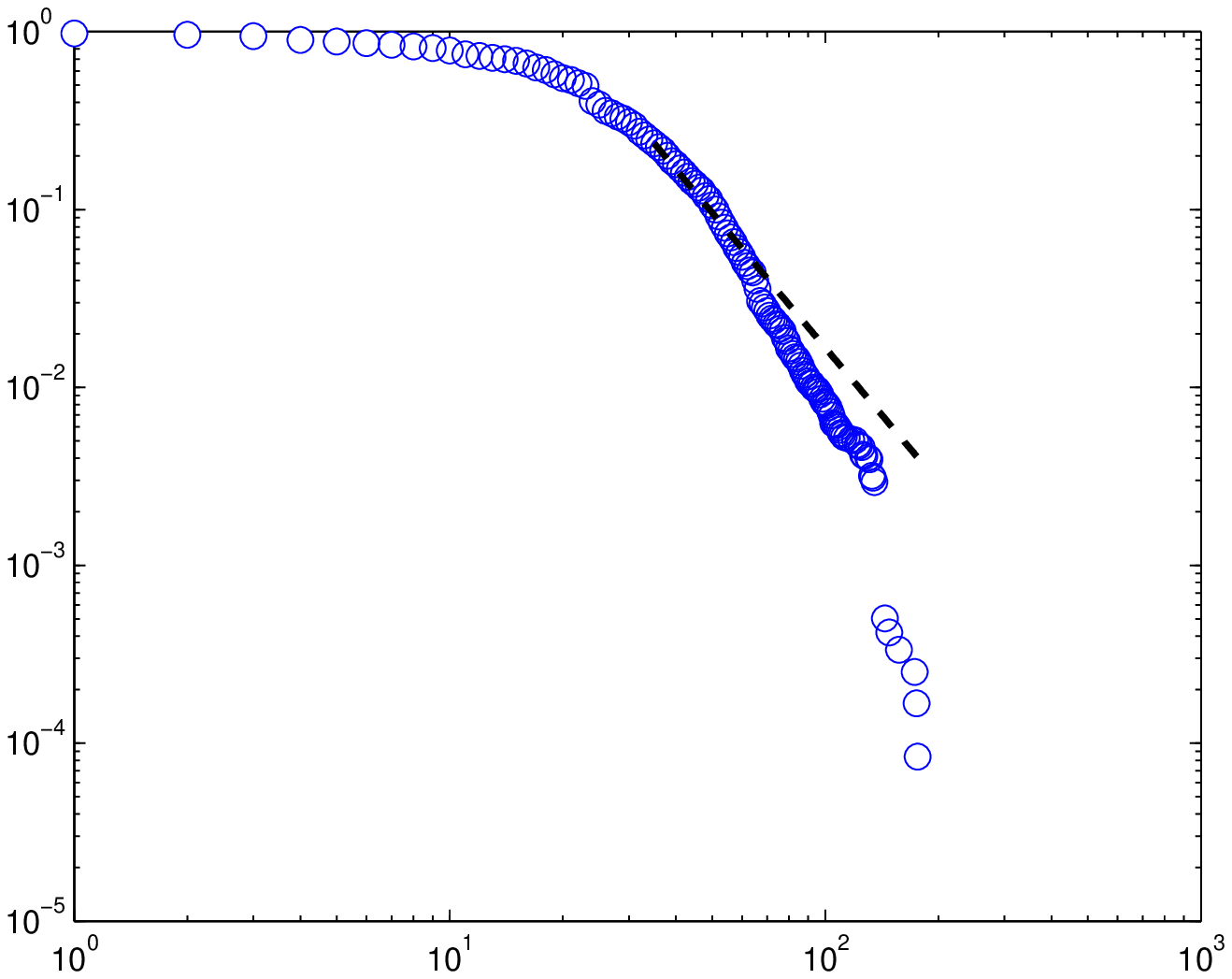}&
  \includegraphics[width=0.45\textwidth]{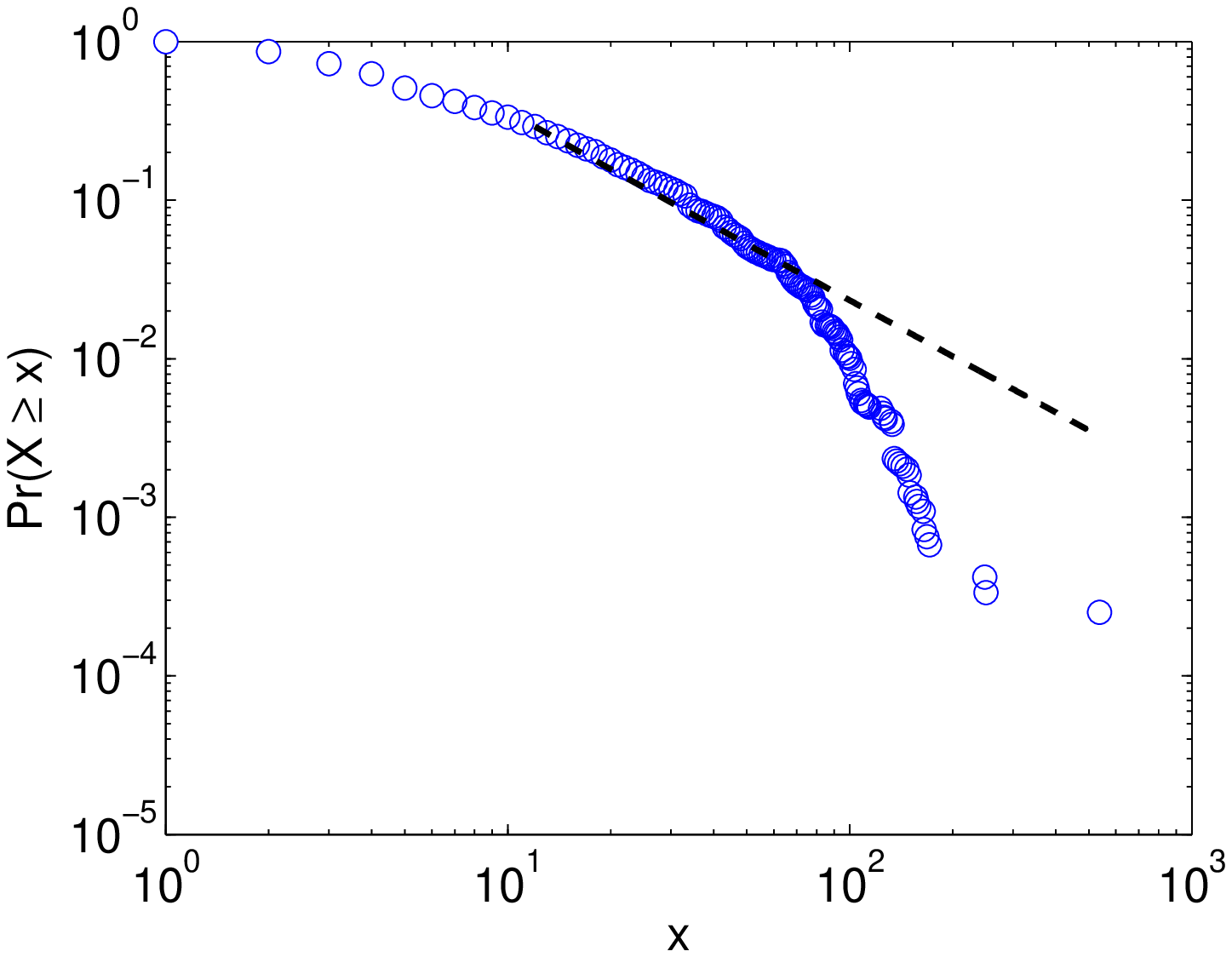}\\
\end{tabular}
\begin{tabular}{c c}
  \includegraphics[width=0.45\textwidth]{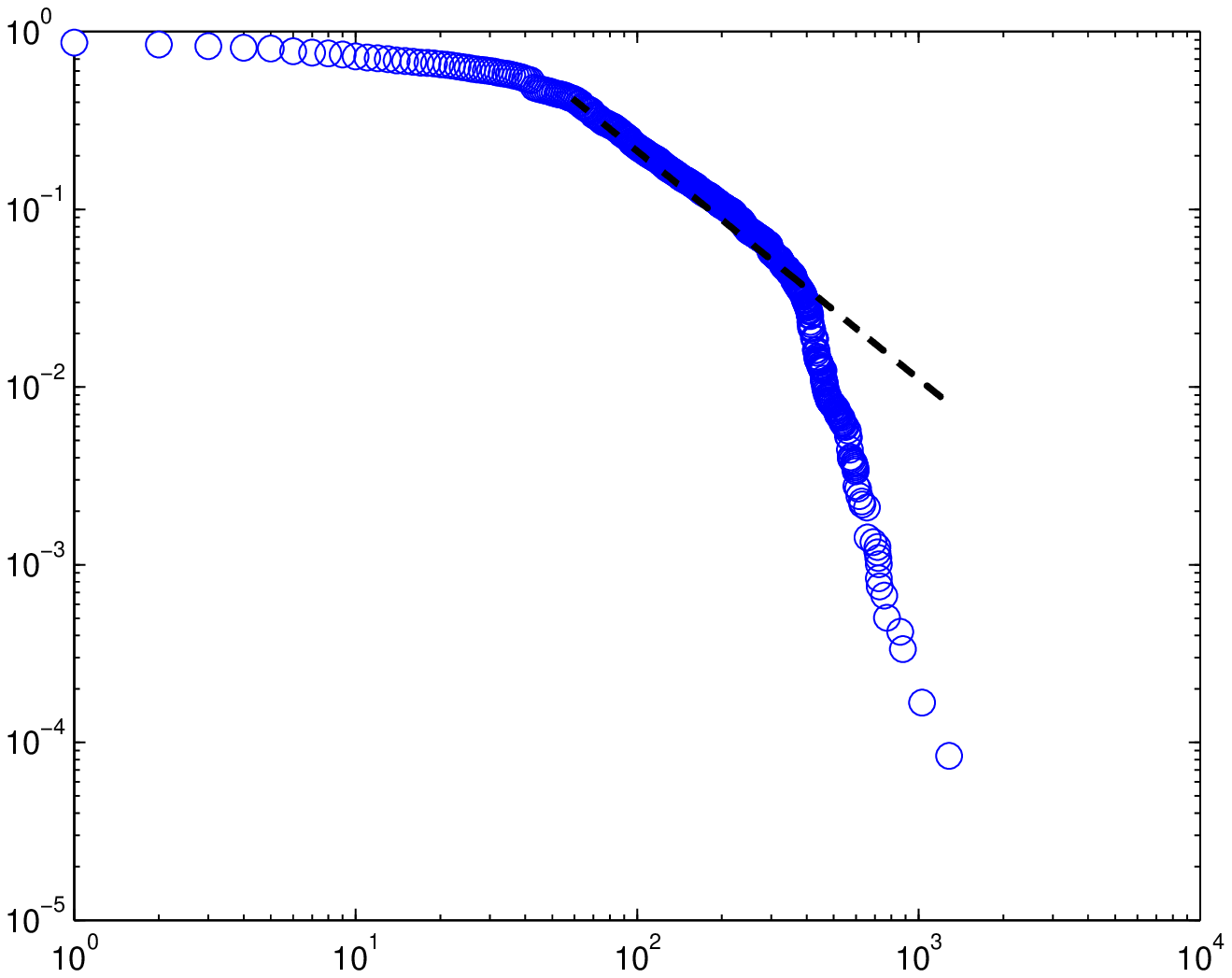}&
  \includegraphics[width=0.45\textwidth]{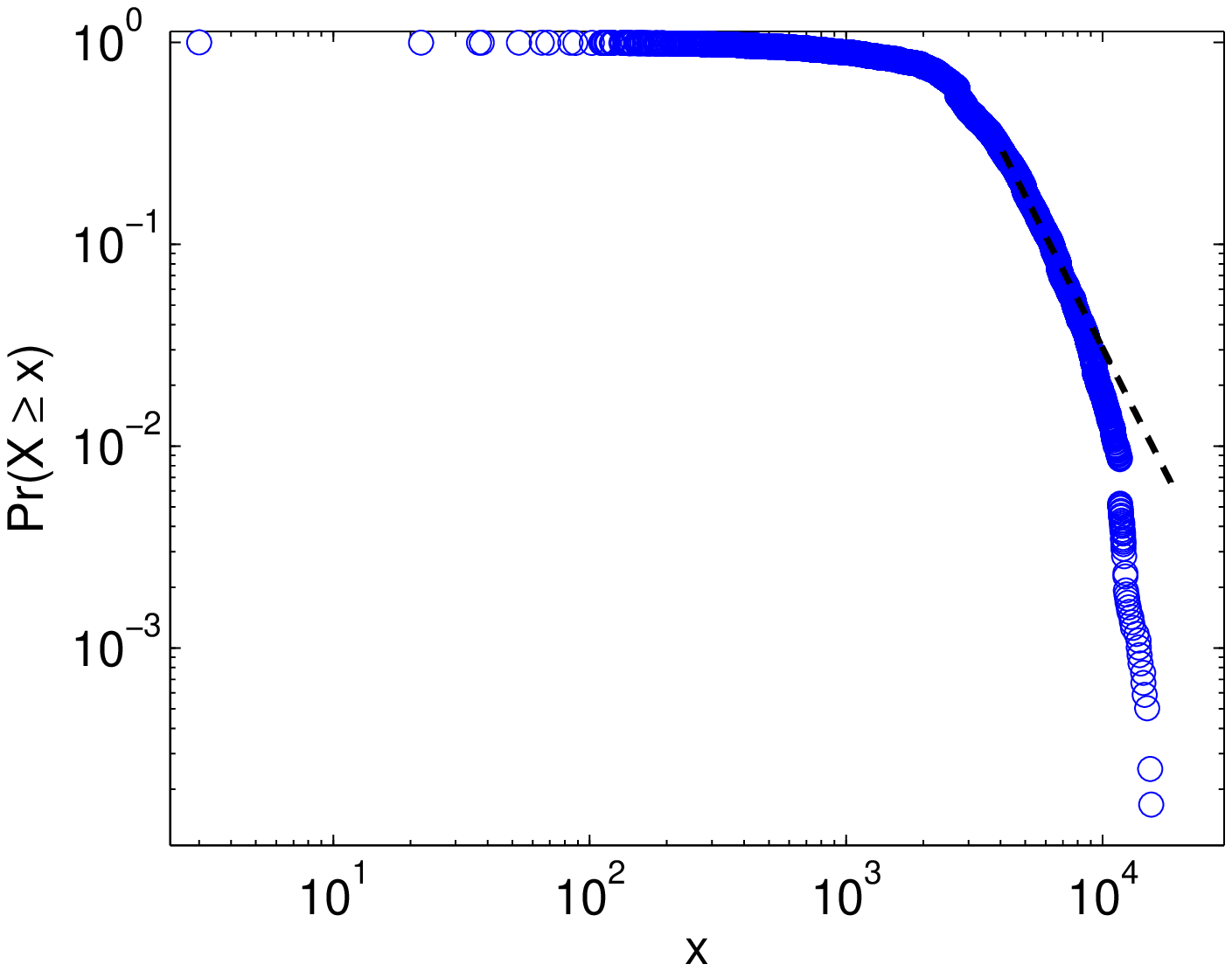}\\
\end{tabular}
\begin{tabular}{c c}
  \includegraphics[width=0.45\textwidth]{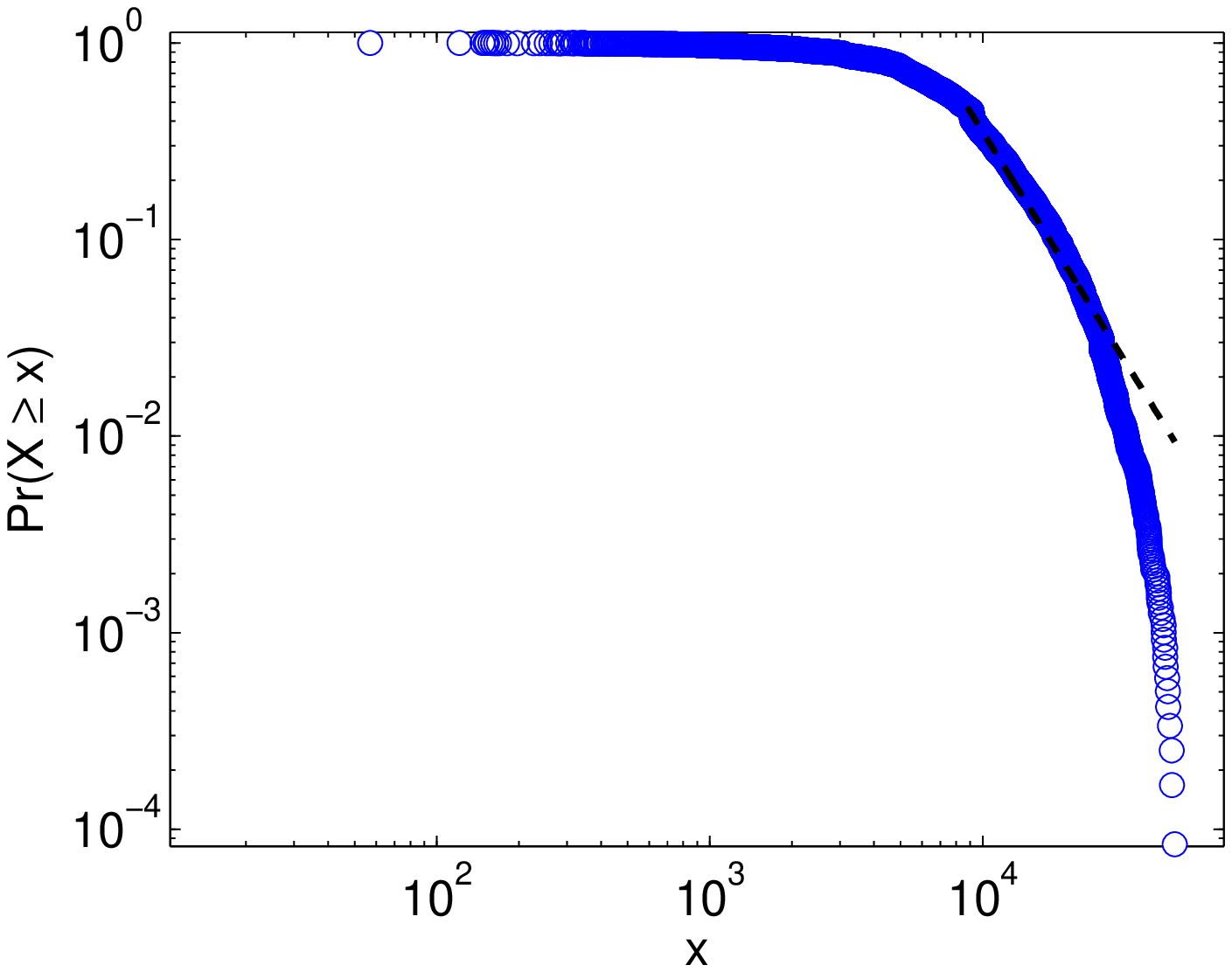}&
  \includegraphics[width=0.45\textwidth]{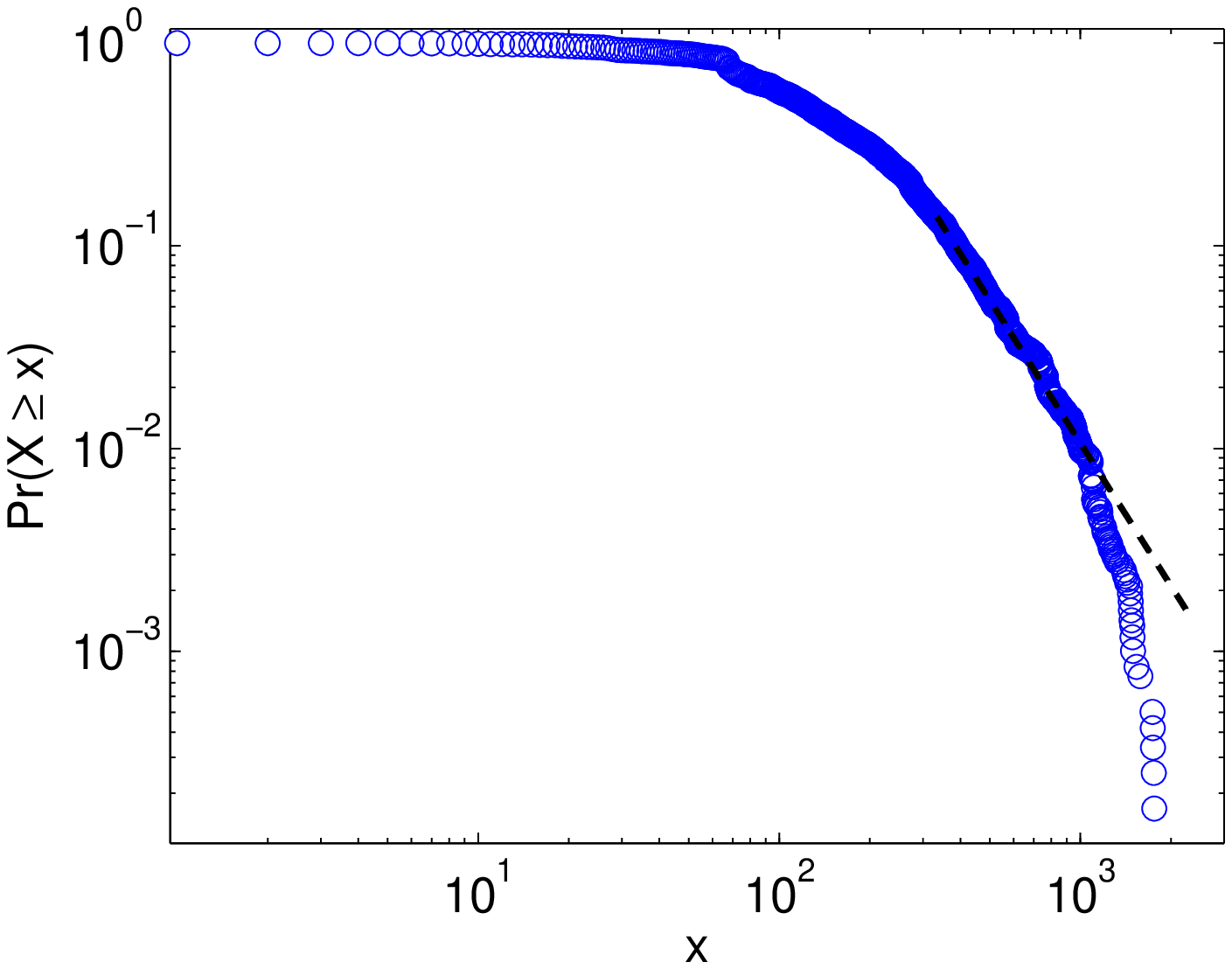}\\
\end{tabular}
\caption{\label{fig:pl3}\textit{Power law best fitting of the metrics Address, Cyclomatic, Comments, ABI, Bytecode and LOCS.}}
\end{figure}

Fig. \ref{fig:pl3} finally shows that a good candidate for a power law in the tail is the LOC metric, 
supported by a KS coefficient of significance of about 0.039. This suggests that also for the Smart Contract code 
the main size metric in software, the lines of code, shows properties similar to those of standard software systems. 
Also the Address metric displays a reasonable power law regime for a range of its values, showing a behaviour
similar to that found for the metric ``Name of Variables'' in Java software \cite{Concas:2007}. Thus the usage of the keyword 
``Address'' in Smart Contracts occurs in quantities which remind the usage of variable names in Java.  

Finally we analyzed all the statistical distributions using a log-normal best fitting model. 

\begin{figure}[!ht]
\begin{tabular}{c c}
  \includegraphics[width=0.45\textwidth]{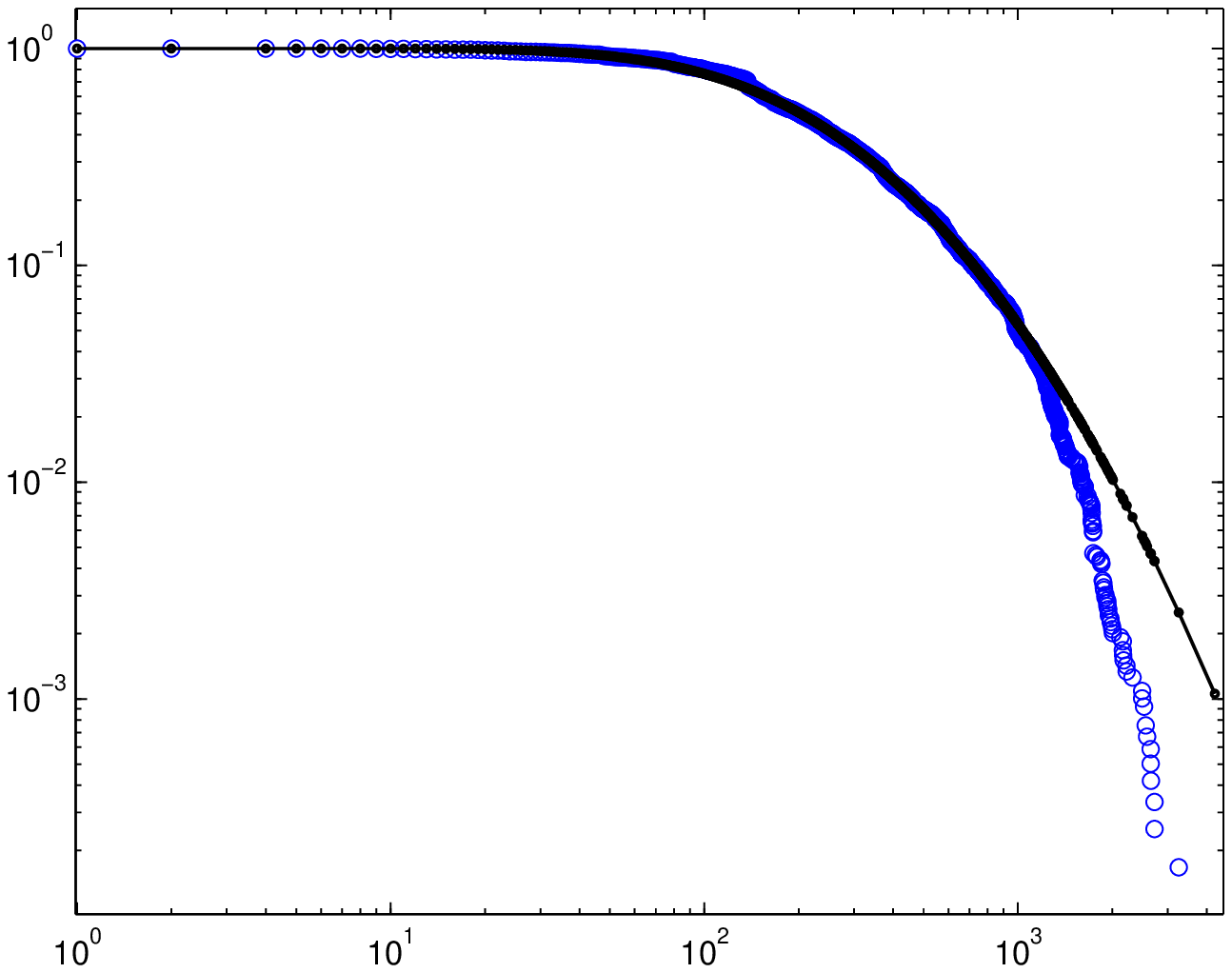}&
  \includegraphics[width=0.45\textwidth]{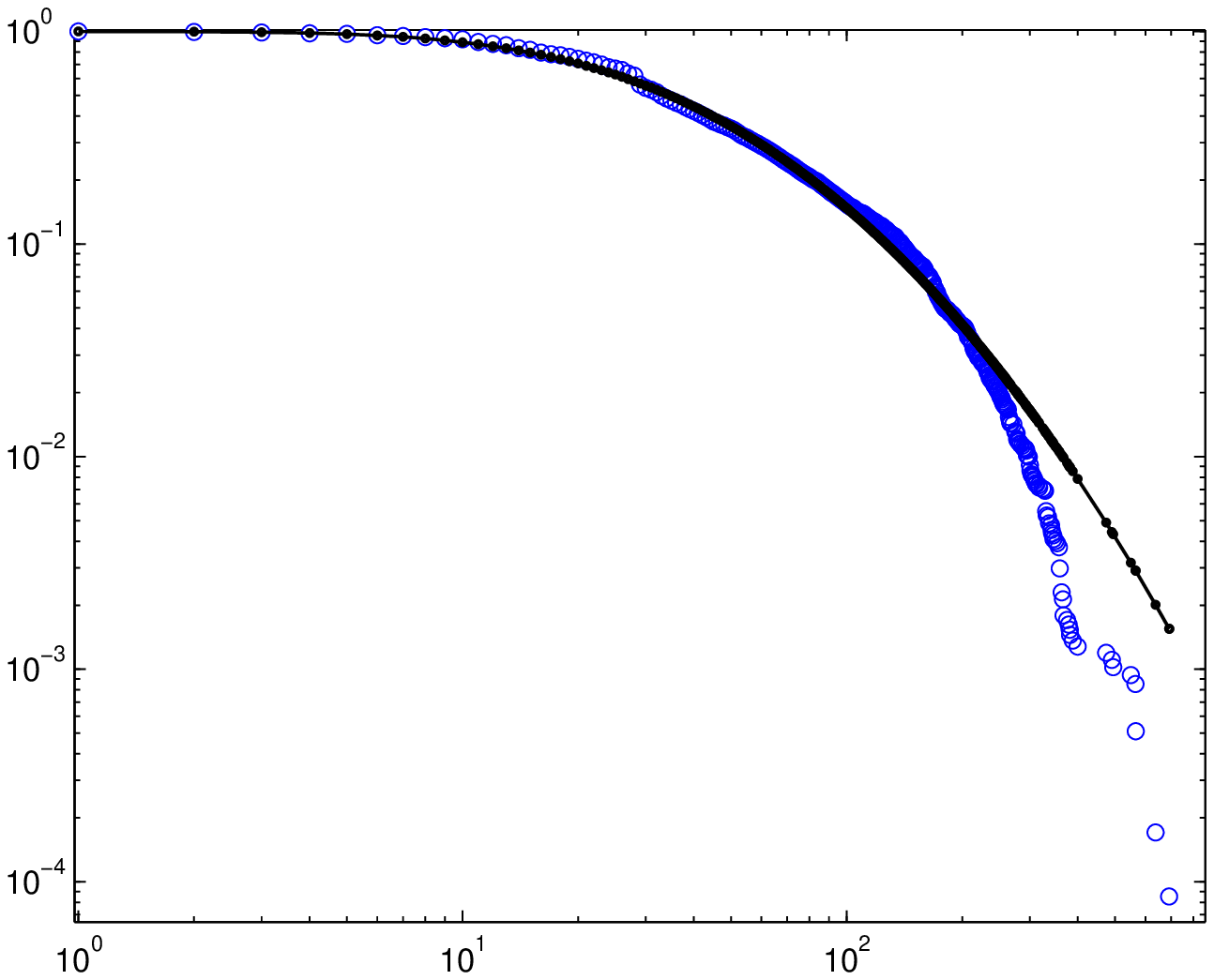} \\
\end{tabular}
\begin{tabular}{c c}
  \includegraphics[width=0.45\textwidth]{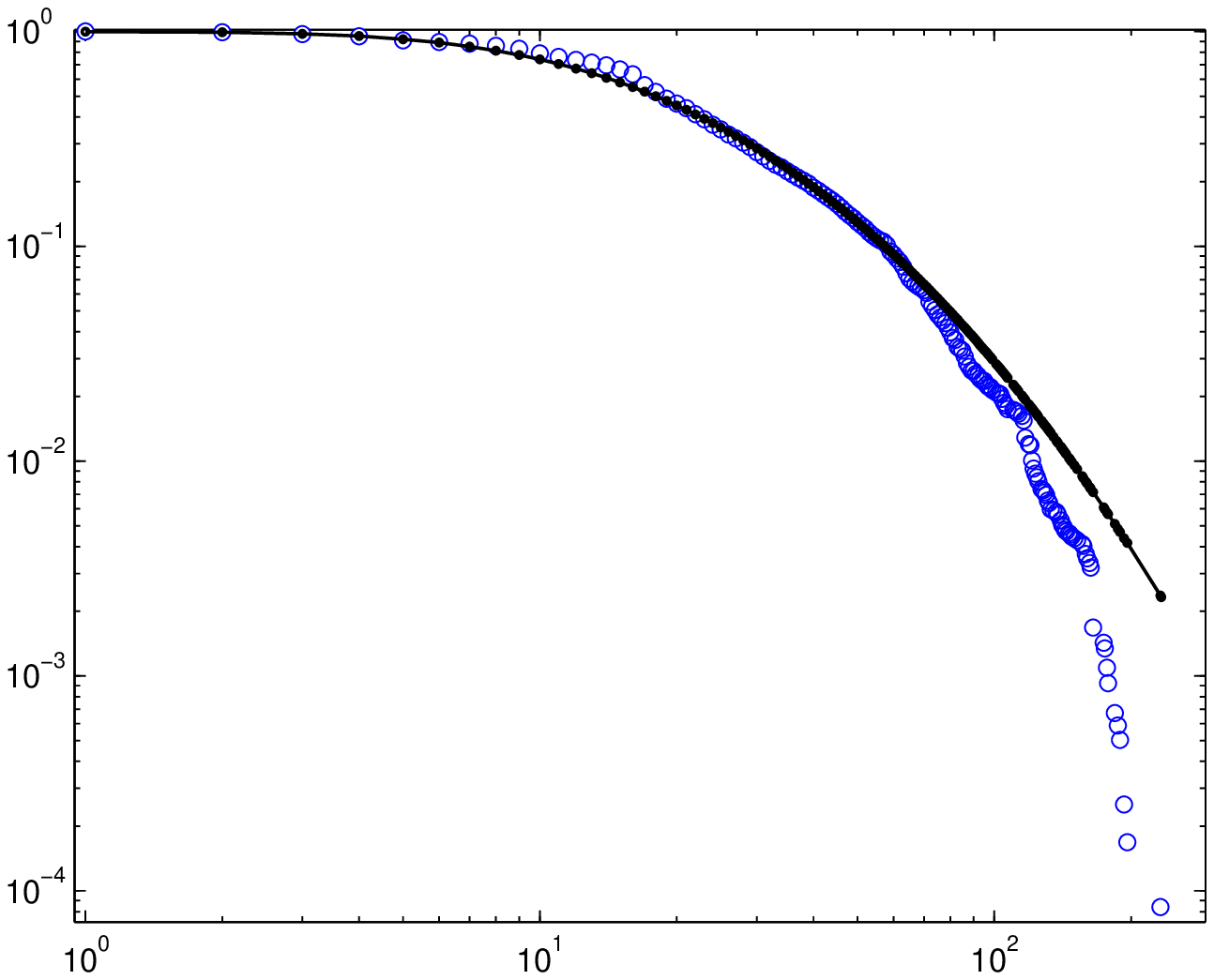}&
  \includegraphics[width=0.45\textwidth]{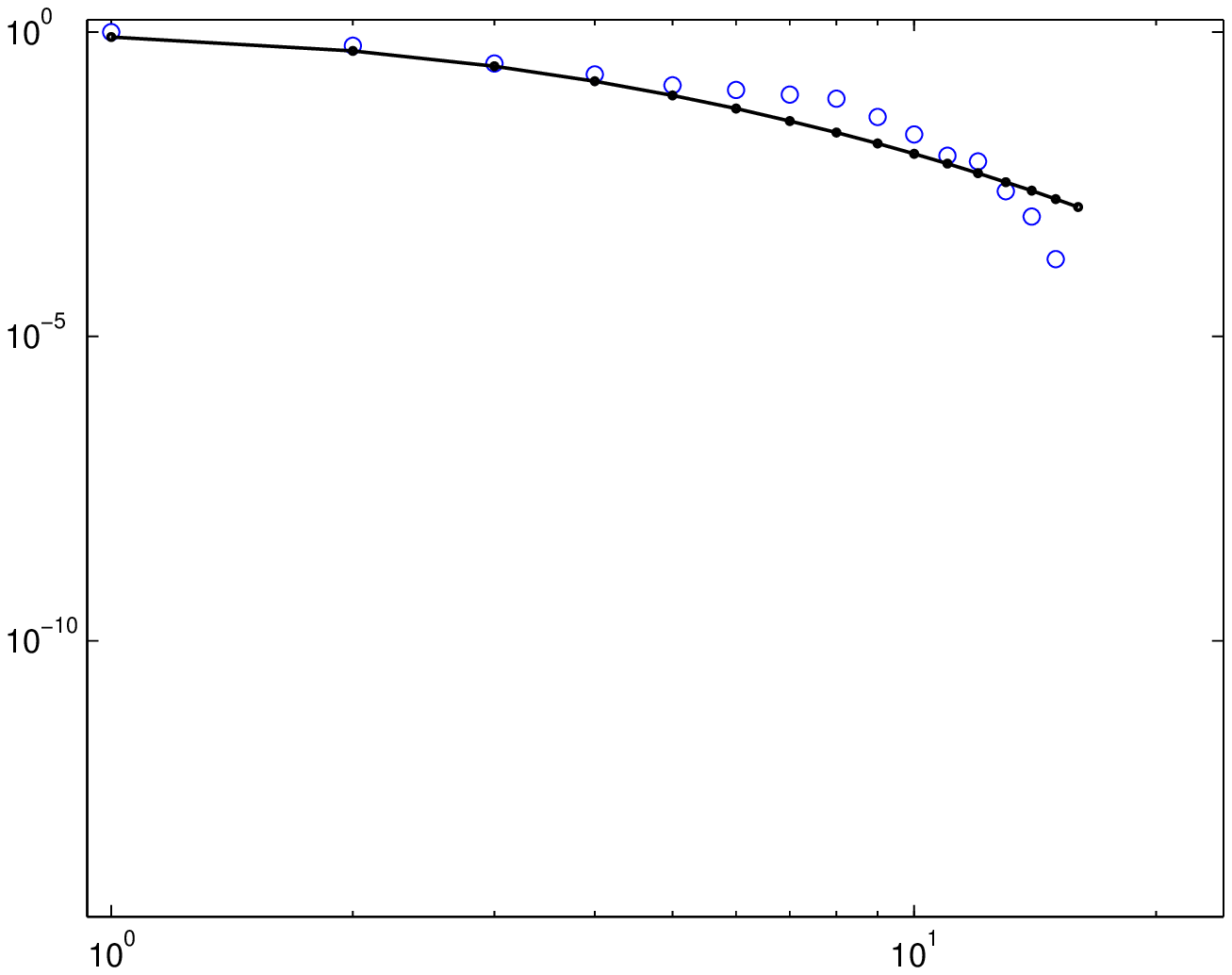}\\
\end{tabular}
\caption{\label{fig:ln1}\textit{Log Normal best fitting of the metrics Total lines, Blanks, Function and Payable}}
\end{figure}

In Fig. \ref{fig:ln1} we show the Lognormal best fitting curves together with the empirical cumulative 
distribution functions for the Smart Contracts metrics Total lines, Blanks, Function and Payable.
The first three metrics are nicely fitted by the Lognormal statistical distribution in the bulk, for low 
values of the metrics, but not in the tail, even if the $R^2$ is quite close to one for each case ($R^2 \geq 0.95$). 
Such result confirms the previouos one obtained for the power law model \cite{Newmann2}. The best fitting lacks mainly in the 
tail of the distribution, as expected. In fact the empirical distribution drops more rapidly than the best fitting 
curve because of the cut-off for large values of the metrics. This may be explained by the hypothesis that 
Smart Contract size metrics, like Total Lines of code, Functions and Blanks are upper bounded according to 
the size costraints associated to the deployment of Smart Contracts into the blockchain. The Payable metric results
in a too poor statistic to be well fitted by a Lognormal distribution.

\begin{figure}[!ht]
\begin{tabular}{c c}
  \includegraphics[width=0.45\textwidth]{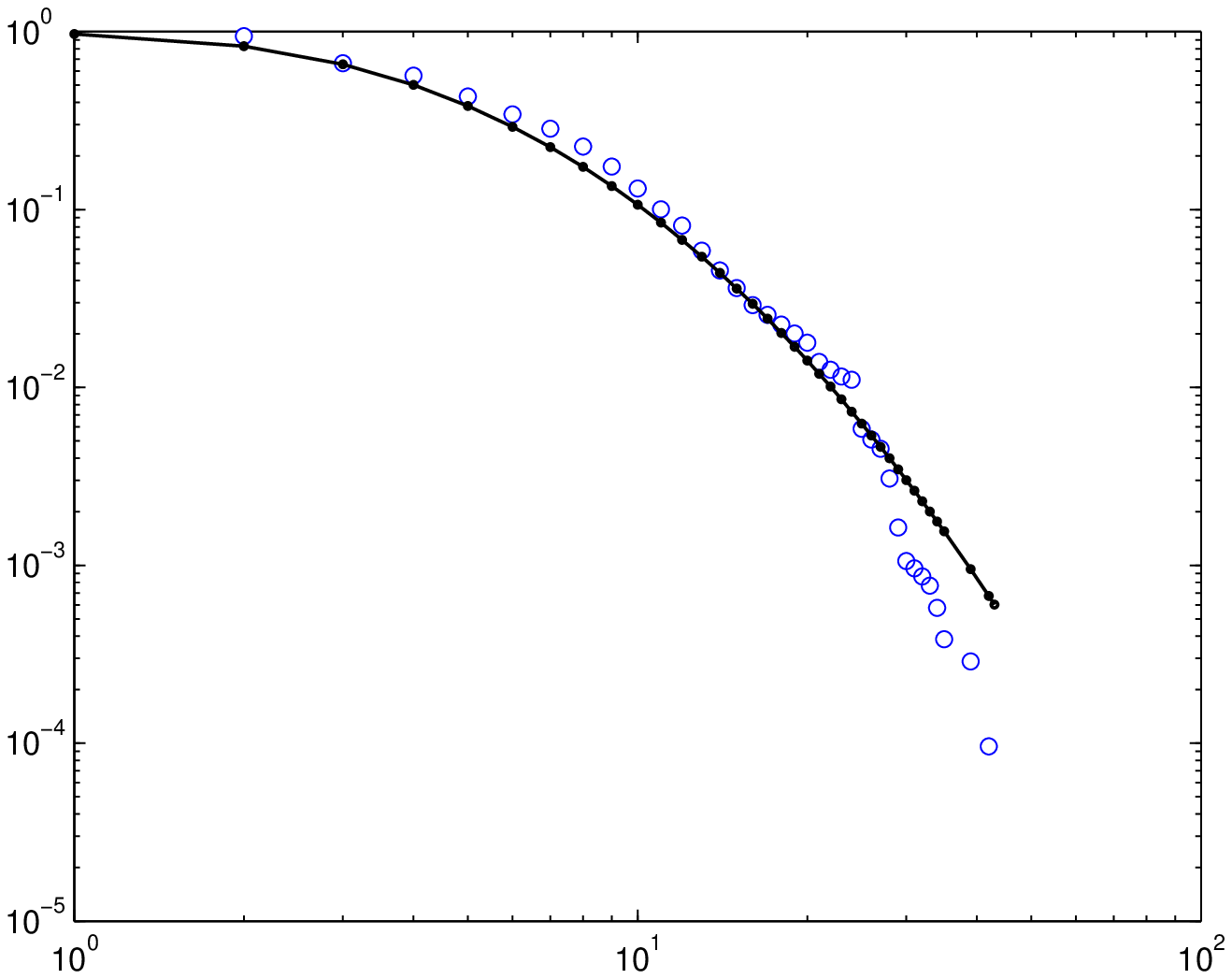}&
  \includegraphics[width=0.45\textwidth]{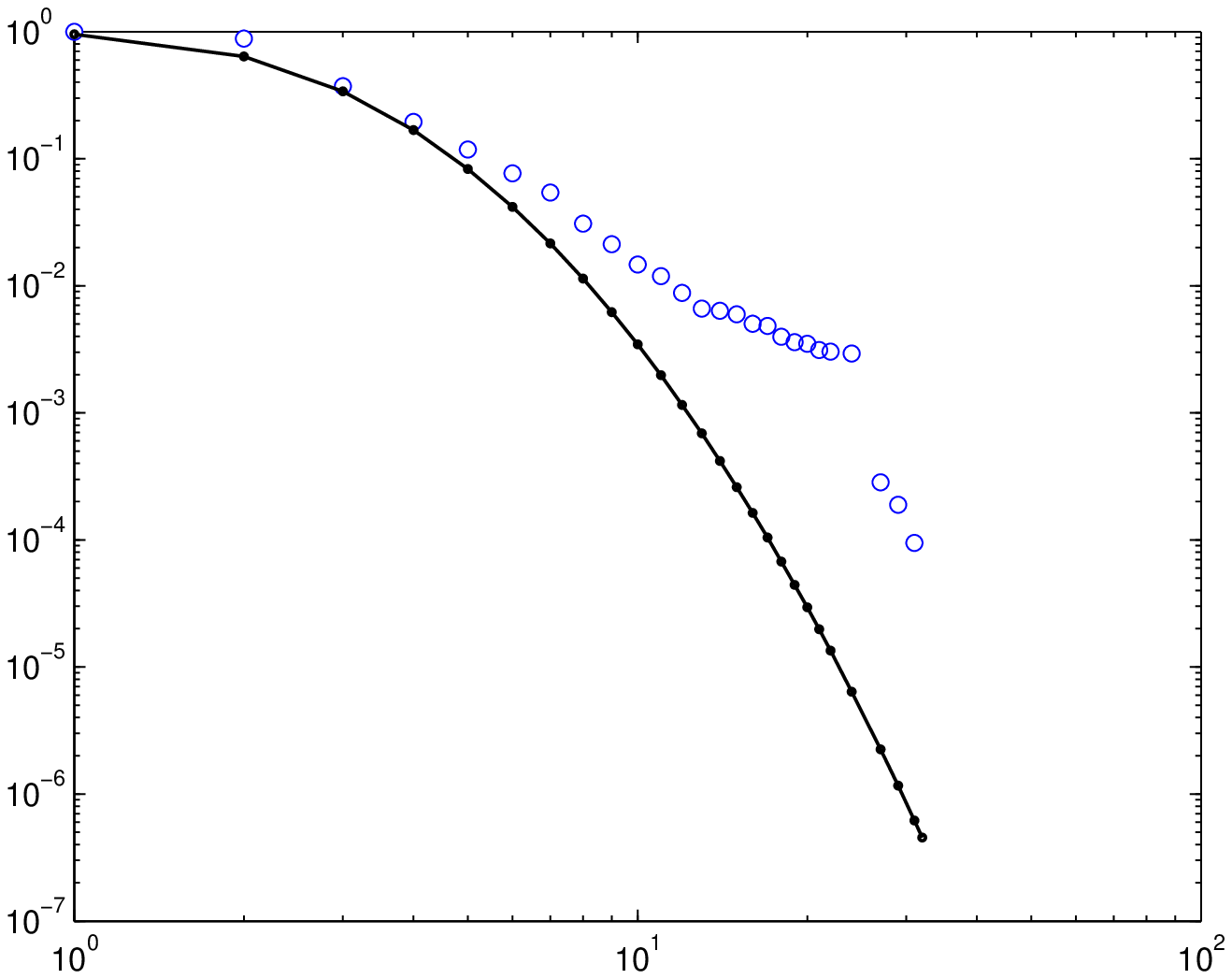} \\
\end{tabular}
\begin{tabular}{c c}
  \includegraphics[width=0.45\textwidth]{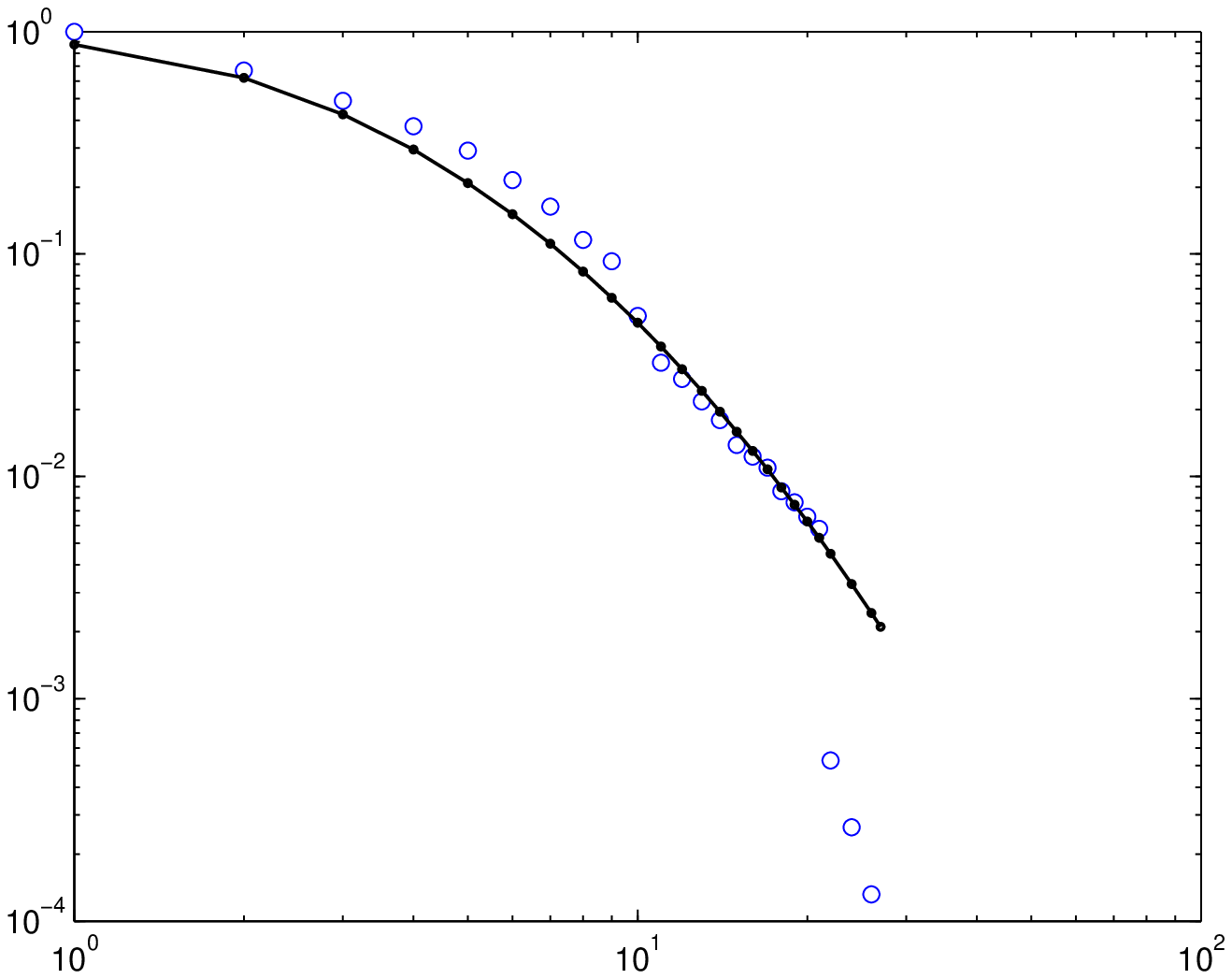}&
  \includegraphics[width=0.45\textwidth]{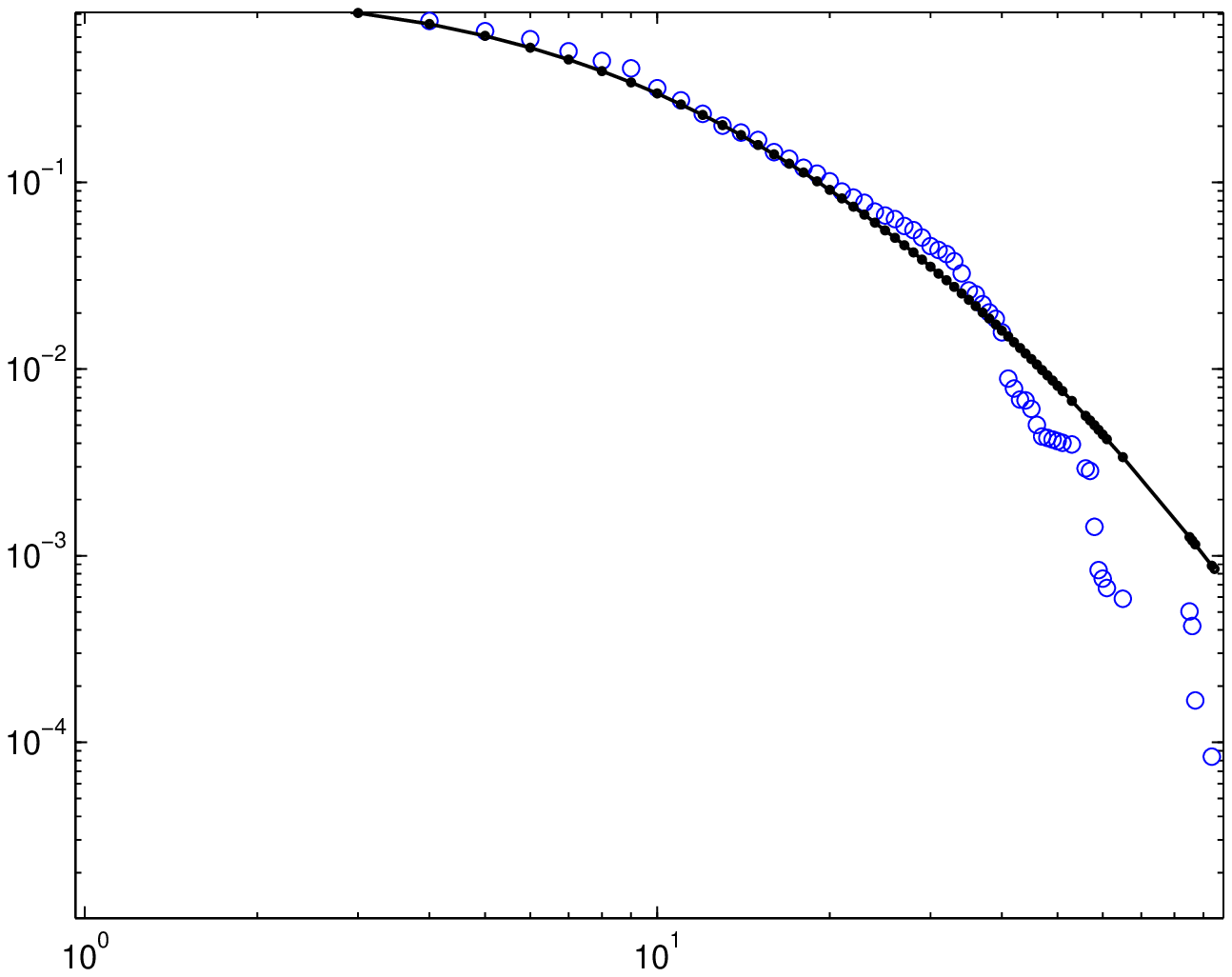}\\
\end{tabular}
\caption{\label{fig:ln2}\textit{Log Normal best fitting of the metrics Events, Mapping, Modifier and Contract.}}
\end{figure}

Fig. \ref{fig:ln2} show the metrics Events, Mapping, Modifier and Contract. Mapping cannot be well fitted 
by a Lognormal, as it was vety well explained by a power law in the range corresponding to the bulk of the 
distribution rather than in the tail. 
Also Events and Modifier do not suite a Lognormal distribution and their $R^2$ values are lower than 0.95. 
Finally Contract is quite well approximated in the bulk, but not in the tail, confirming once again the power law 
best fitting results.

\begin{figure}[!ht]
\begin{tabular}{c c}
  \includegraphics[width=0.45\textwidth]{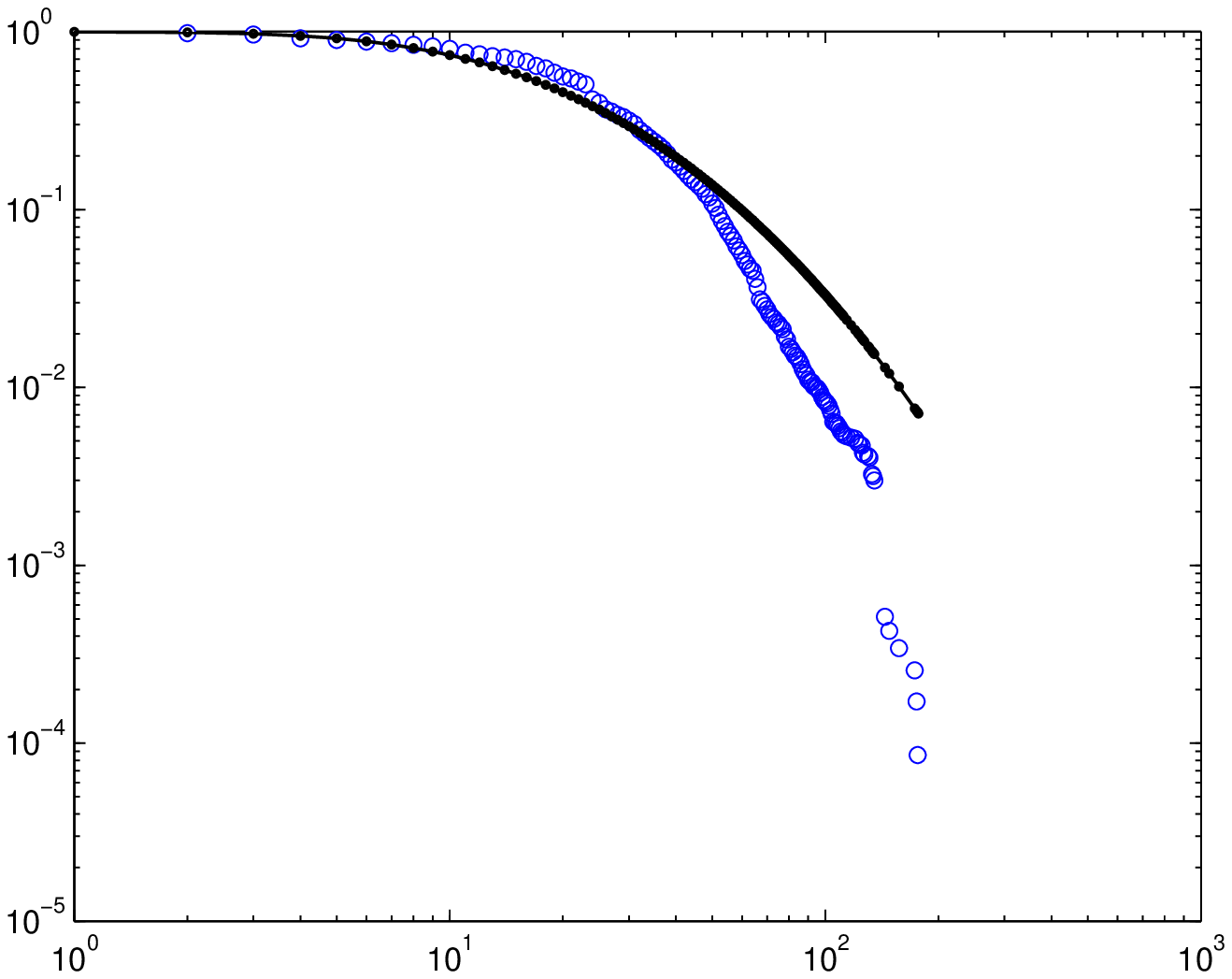}&
  \includegraphics[width=0.45\textwidth]{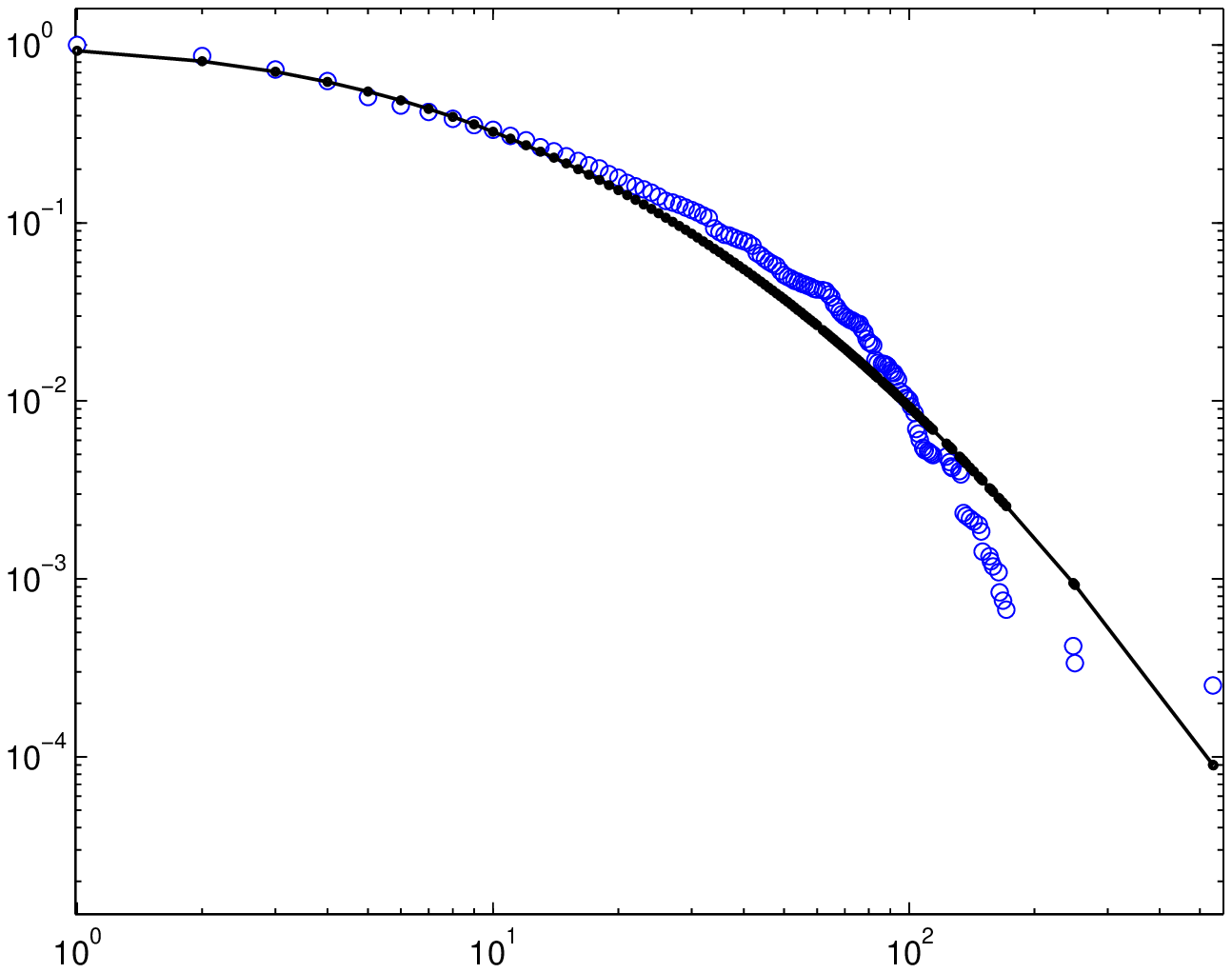}\\
\end{tabular}
\begin{tabular}{c c}
  \includegraphics[width=0.45\textwidth]{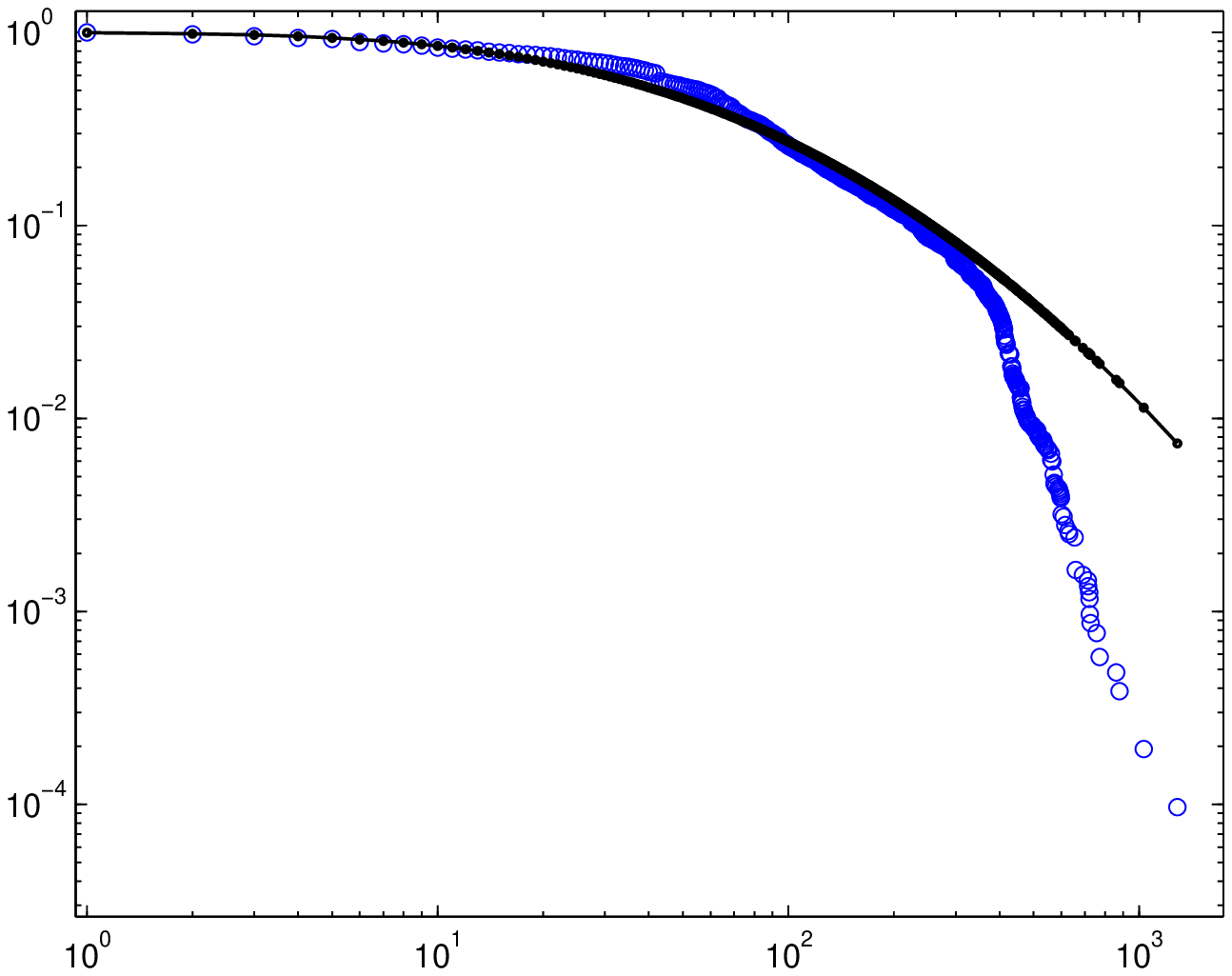}&
  \includegraphics[width=0.45\textwidth]{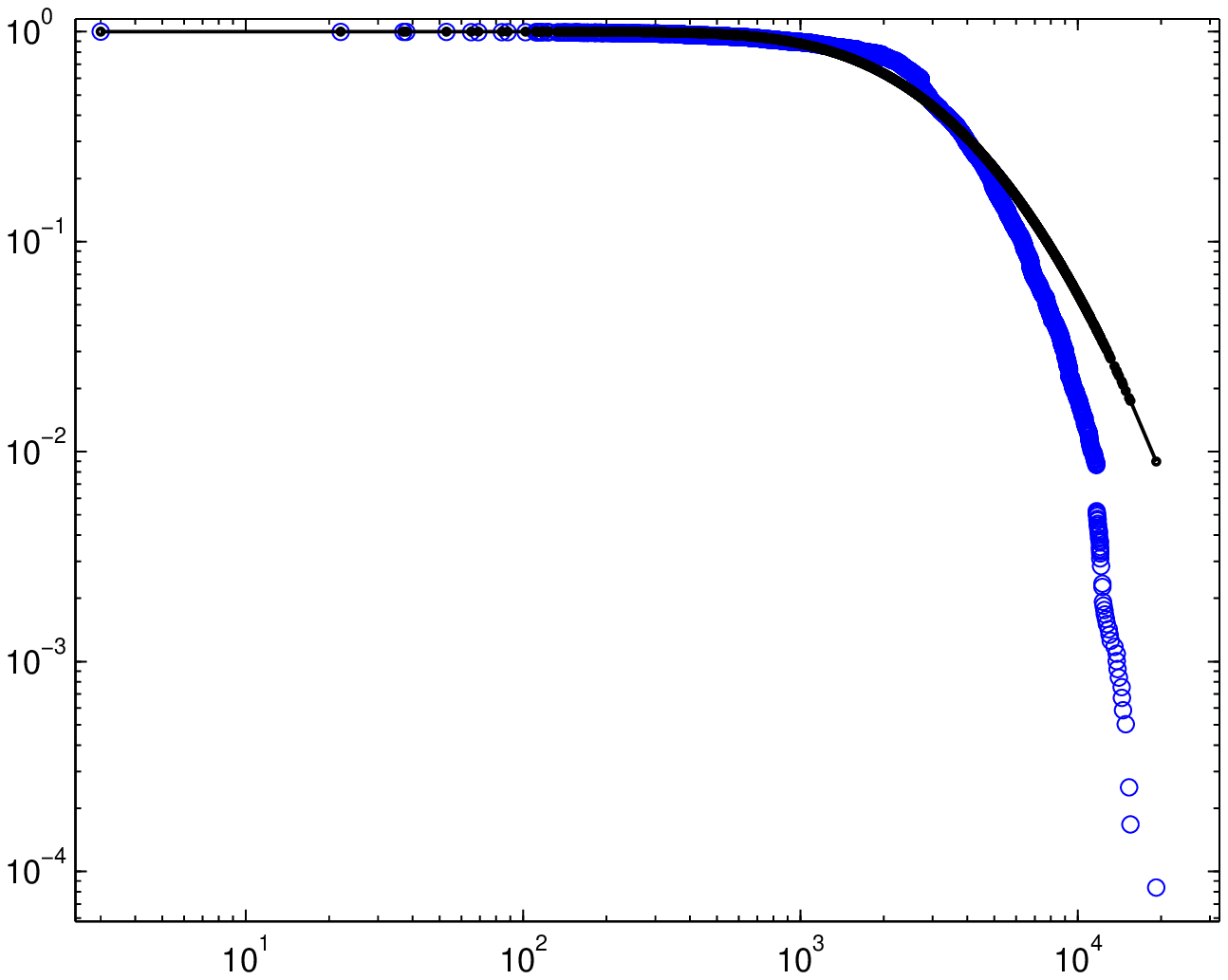}\\
\end{tabular}
\begin{tabular}{c c}
  \includegraphics[width=0.45\textwidth]{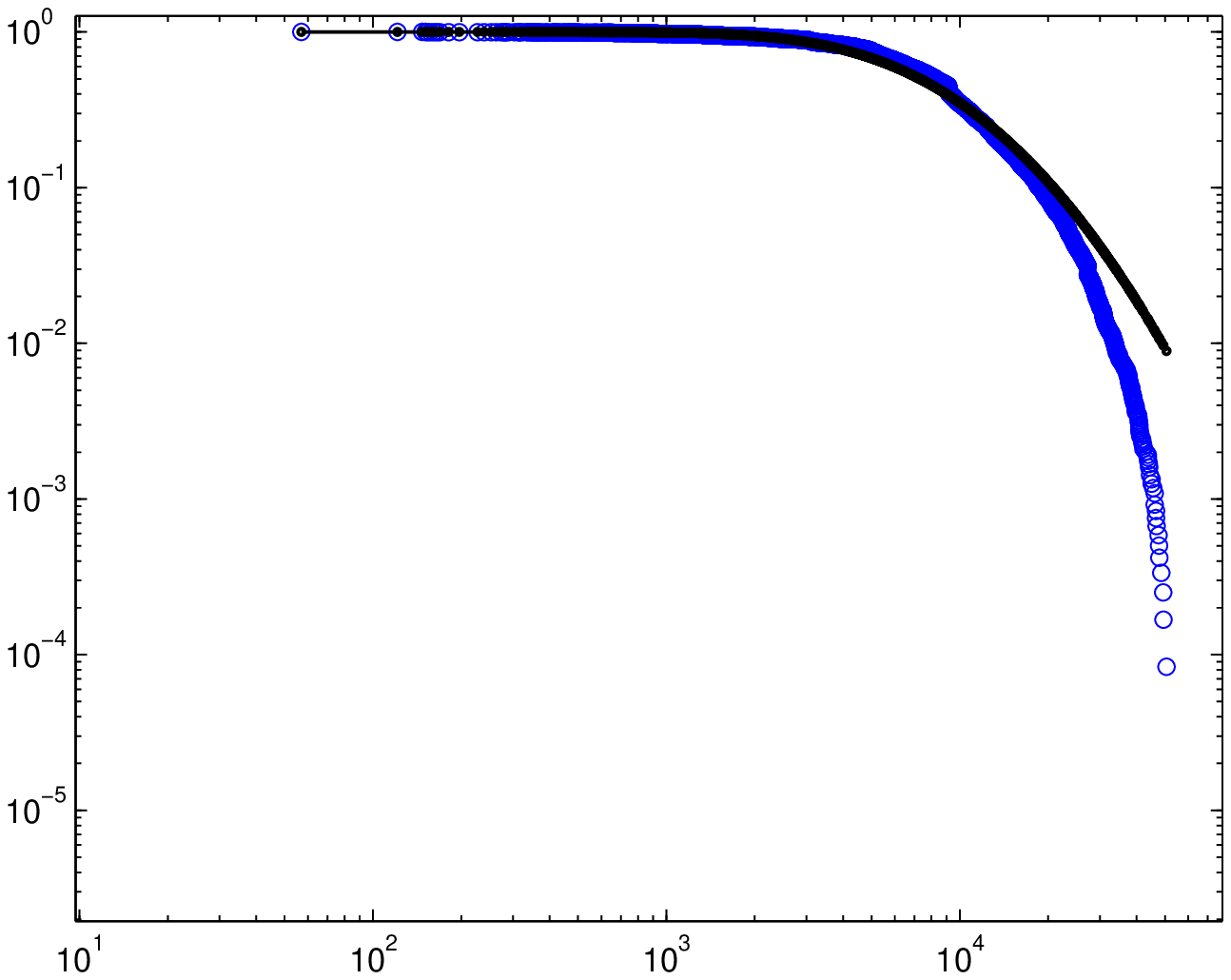}&
  \includegraphics[width=0.45\textwidth]{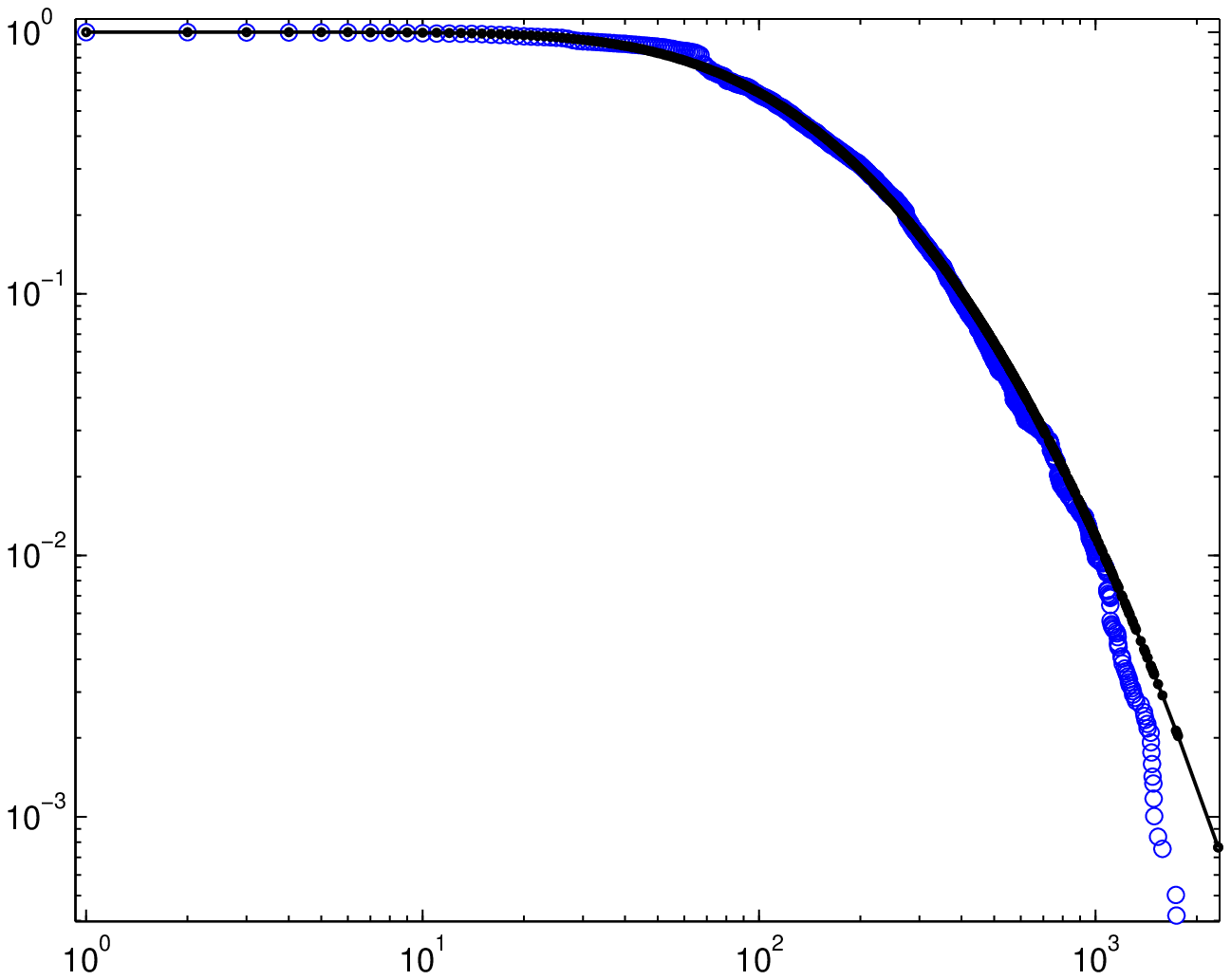}\\
\end{tabular}
\caption{\label{fig:ln3}\textit{Log Normal best fitting of the metrics Address, Cyclomatic, Comments, ABI, Bytecode and LOCS.}}
\end{figure}

Finally Fig. \ref{fig:ln3} shows that the initial parts of Bytecode and ABI metrics well overlap with the Lognormal
but as soon as the values crosses the central ones observed in the corresponding histograms the Lognormal
curves tend to miss the empirical ones which drops quickly and do not display power law in the tail. 

Address, Cyclomatic ad Comments rapidly drop with respect to the Lognormal model, even if the initial 
part presents some overlap with it. Again this may be ascribed to the upper bounds which limit the 
range of values reachable by these metrics. In particular Comments are less, on average, than in traditional software 
development. This is maybe due to the fact that Smart Contrac software code is written with specific purpose and 
contraints, so that the same patterns are most likely found and do not need comment lines. 

Finally the LOC metric is quite well represented by the Lognormal distribution both on the bulk and in the tail, 
and presents an $R^2$ value larger than 0.98. This is quite in agreement with the results found in literature 
for the LOC metric in traditional software systems \cite{Concas:2006}. In some sense this result is different from the others 
since it seems that this metric is not influenced by the peculiarity that can belong to Smart Contract software
and tends to preserve the same statistical features found in traditional software systems.

\section{Conclusions}\label{Conclusions}

In this paper we studied Smart Contracts software metrics extracted from a data set 
of more than 12000 Smart Contracts deployed on the Ethereum blockchain. 
We were interested in determining if, given the peculiarity related Smart Contract
software development, the corresponding software metrics display differences
in their statistical properties with respect to metrics extracted from traditional 
software systems and already largely studied in literature. 

The assumptions are that resources are limited on the blockchain and such limitations
may influence the way Smart Contracts are written. Our analisys dealt with source code
metrics as well as with ABI and bytecode of Smart Contracts. 

Our main results show that, overall, the exposure of Smart Contracts to the interaction 
with the blockchain as qualitatively measured in terms of ABI size are quite similar 
to each other and there are not outlyers Contracts. The distribution is compatible with 
a bell shaped statistical distribution where most of values tend to lie around 
a central value with some dispersion around it.

In general Smart Contracts metrics tend to suffer from blockchain limited resources 
constraints, since they tend to assume limited upper values. There is not the 
ubiquitous presence of fat tail distributions where there are values very far from the mean, 
even order of magnitude larger, as typical in traditional software. In Smart Contract 
software metrics large variations from the mean are substantially unknown and all the values
are generally inta a range of few standard deviations from the mean. 

Fianally the Smart Contract lines of code is the metric which more closely follow the 
statistical ditribution of the corresponding metric in traditional software system 
and shows a truncated power law in the tail and an overall distribution which is well 
explained by a Lognormal distribution.

\end{document}